\documentclass[twocolumn,english,draftclsnofoot, 12pt]{IEEEtran}
\usepackage[T1]{fontenc}
\usepackage{color}
\usepackage{babel}
\usepackage{float}
\usepackage{amsmath}
\usepackage{amssymb}
\usepackage{graphicx}
\usepackage[unicode=true,
 bookmarks=true,bookmarksnumbered=true,bookmarksopen=true,bookmarksopenlevel=1,
 breaklinks=false,pdfborder={0 0 0},backref=false,colorlinks=false]
 {hyperref}
\hypersetup{pdftitle={Your Title},
 pdfauthor={Your Name},
 pdfpagelayout=OneColumn, pdfnewwindow=true, pdfstartview=XYZ, plainpages=false}

\makeatletter

\floatstyle{ruled}
\newfloat{algorithm}{tbp}{loa}
\providecommand{\algorithmname}{Algorithm}
\floatname{algorithm}{\protect\algorithmname}

\ifCLASSOPTIONcompsoc
\usepackage[caption=false,font=normalsize,labelfont=sf,textfont=sf]{subfig}
\else
\usepackage[caption=false,font=footnotesize]{subfig}
\fi

\onecolumn
\usepackage{algorithm}
\usepackage{algorithmic}

\newtheorem{assumption}{Assumption}
\newtheorem{remark}{Remark}
\newtheorem{lemma}{Lemma}
\newtheorem{theorem}{Theorem}
\newtheorem{definition}{Definition}
\newtheorem{problem}{Problem}

\makeatother

\begin{document}

\title{Delay-Aware Uplink Fronthaul Allocation in Cloud Radio Access Networks}

\author{Wei Wang,~\IEEEmembership{Member,~IEEE,} Vincent K. N. Lau,~\IEEEmembership{Fellow,~IEEE,}
\\Mugen Peng,~\IEEEmembership{Senior Member,~IEEE}%
\thanks{This work is supported in part by National Natural Science Foundation
of China (No. 61261130585), Research Grants Council (RGC) of Hong
Kong (No. N\_HKUST605/13), and Hong Kong Scholars Program (No. 2012T50566).%
}%
\thanks{W. Wang and V. K. N. Lau are with Department of Electrical and Computer
Engineering, Hong Kong University of Science and Technology, Clear
Water Bay, Hong Kong. Email: \{eewangw,eeknlau\}@ust.hk%
}%
\thanks{M. Peng is with School of Information and Communication Engineering,
Beijing University of Posts and Telecommunications, Beijing, China.
Email: pmg@bupt.edu.cn%
}\vspace{-1cm}}
\maketitle
\begin{abstract}
In cloud radio access networks (C-RANs), the baseband units and radio
units of base stations are separated, which requires high-capacity
fronthaul links connecting both parts. In this paper, we consider
the delay-aware fronthaul allocation problem for C-RANs. The stochastic
optimization problem is formulated as an infinite horizon average
cost Markov decision process. To deal with the curse of dimensionality,
we derive a closed-form approximate priority function and the associated
error bound using perturbation analysis. Based on the closed-form
approximate priority function, we propose a low-complexity delay-aware
fronthaul allocation algorithm solving the per-stage optimization
problem. The proposed solution is further shown to be asymptotically
optimal for sufficiently small cross link path gains. Finally, the
proposed fronthaul allocation algorithm is compared with various baselines
through simulations, and it is shown that significant performance
gain can be achieved.\end{abstract}
\begin{IEEEkeywords}
cloud radio access networks, fronthaul link, delay optimization, perturbation
analysis, Markov decision process
\end{IEEEkeywords}
\newpage

\section{Introduction}

The cloud radio access network (C-RAN) \cite{CRAN} provides a new
architecture for 5G cellular systems. In C-RANs, the baseband processing
of base stations is carried out in the cloud, i.e., a centralized
base band unit (BBU), which launches joint signal processing with
coordinated multi-point transmission (CoMP) and makes it possible
to mitigate inter-cell interference. The separation of the BBU and
the radio units (RUs) brings a new segment, i.e., fronthaul links,
to connect both parts. The limited capacities of fronthaul links have
a significant influence on the system performance of C-RANs.

There are several existing works on fronthaul links in C-RANs. Efficient
signal quantization/compression for fronthaul links is designed to
maximize the network throughput for the uplink and downlink in \cite{compression}
and \cite{compression2}, respectively. In \cite{joint}, fronthaul
quantization and transmit power control are optimized jointly. In
\cite{comp}, energy-efficient CoMP is designed for downlink transmission
considering fronthaul capacity. In \cite{WZcompression}, the capacities
of fronthaul links are allocated under a sum capacity constraint to
maximize the total throughput. In \cite{mobicom}, the fronthaul links
are reconfigured to apply appropriate transmission strategies in different
parts according to both heterogeneous user profiles and dynamic traffic
load patterns. However, these existing works have all focused on the
physical layer performance without consideration of bursty data arrivals
at the transmitters or of the delay requirement of the information
flows. Since real-life applications (such as video streaming, web
browsing or VoIP) are delay-sensitive, it is important to optimize
the delay performance of C-RANs.

To take the queueing delay into consideration, the fronthaul allocation
policy should be a function of both the channel state information
(CSI) and the queue state information (QSI). This is because the CSI
reveals the instantaneous transmission opportunities at the physical
layer and the QSI reveals the urgency of the data flows. However,
the associated optimization problem is very challenging. A systematic
approach to the delay-aware optimization problem is through a Markov
Decision Process (MDP). In general, the optimal control policy can
be obtained by solving the well-known \emph{Bellman equation}. Conventional
solutions to the Bellman equation, such as brute-force value iteration
or policy iteration \cite{DPcontrol}, have huge complexity (i.e.,
the curse of dimensionality), because solving the Bellman equation
involves solving an exponentially large system of non-linear equations.

In this paper, we focus on minimizing the average delay by fronthaul
allocation. There are two technical challenges associated with the
fronthaul allocation optimization problem:
\begin{itemize}
\item \textbf{Challenges due to the Average Delay Consideration}: Unlike
other works which optimize the physical layer throughput, the optimization
involving average delay performance is fundamentally challenging.
This is because the associated problem belongs to the class of \emph{stochastic
optimization} \cite{learning}, which embraces both \emph{information
theory} (to model the physical layer dynamics) and \emph{queueing
theory} (to model the queue dynamics). A key obstacle to solving the
associated Bellman equation is to obtain the priority function, and
there is no easy and systematic solution in general \cite{DPcontrol}.
\item \textbf{Challenges due to the Coupled Queue Dynamics:} The queues
of data flows are coupled together due to the mutual interference.
The associated stochastic optimization problem is a $K$-dimensional
MDP, where $K$ is the number of data flows. This $K$-dimensional
MDP leads to the curse of dimensionality with complexity exponential
to $K$ for solving the associated Bellman equation. It is highly
nontrivial to obtain a low complexity solution for dynamic fronthaul
allocation in C-RANs.
\end{itemize}

In this paper, we model the fronthaul allocation problem as an infinite
horizon average cost MDP and propose a low-complexity delay-aware
fronthaul allocation algorithm. To overcome the aforementioned technical
challenges, we exploit the specific problem structure that the cross
link path gain is usually weaker than the home cell path gain. Utilizing
the \emph{perturbation analysis} technique, we obtain a closed-form
approximate priority function and the associated error bound. Based
on that, we obtain a low-complexity delay-aware fronthaul allocation
algorithm. The solution is shown to be asymptotically optimal for
sufficiently small cross link path gains. Furthermore, the simulation
results show that the proposed fronthaul allocation achieves significant
delay performance gain over various baseline schemes.

The rest of this paper is organized as follows. In Section II, we
establish the wireless access link, fronthaul link and cloud baseband
processing models as well as the queue dynamics. In Section III, we
formulate the fronthaul allocation problem and derive the associated
optimality conditions. In Section IV, we propose a low-complexity
fronthaul allocation solution. Following this, the delay performance
of the proposed algorithm is evaluated by simulation in Section V.
Finally, conclusions are drawn in Section VI.

\section{System Model}

In this section, we introduce the C-RAN topology and the associated
models of the access link, the fronthaul link and the cloud baseband
processing. Based on the models, we obtain the throughput and the
dynamics of packet queues.

\subsection{C-RAN Topology}

\begin{figure}[t]
\centering\includegraphics[width=5in]{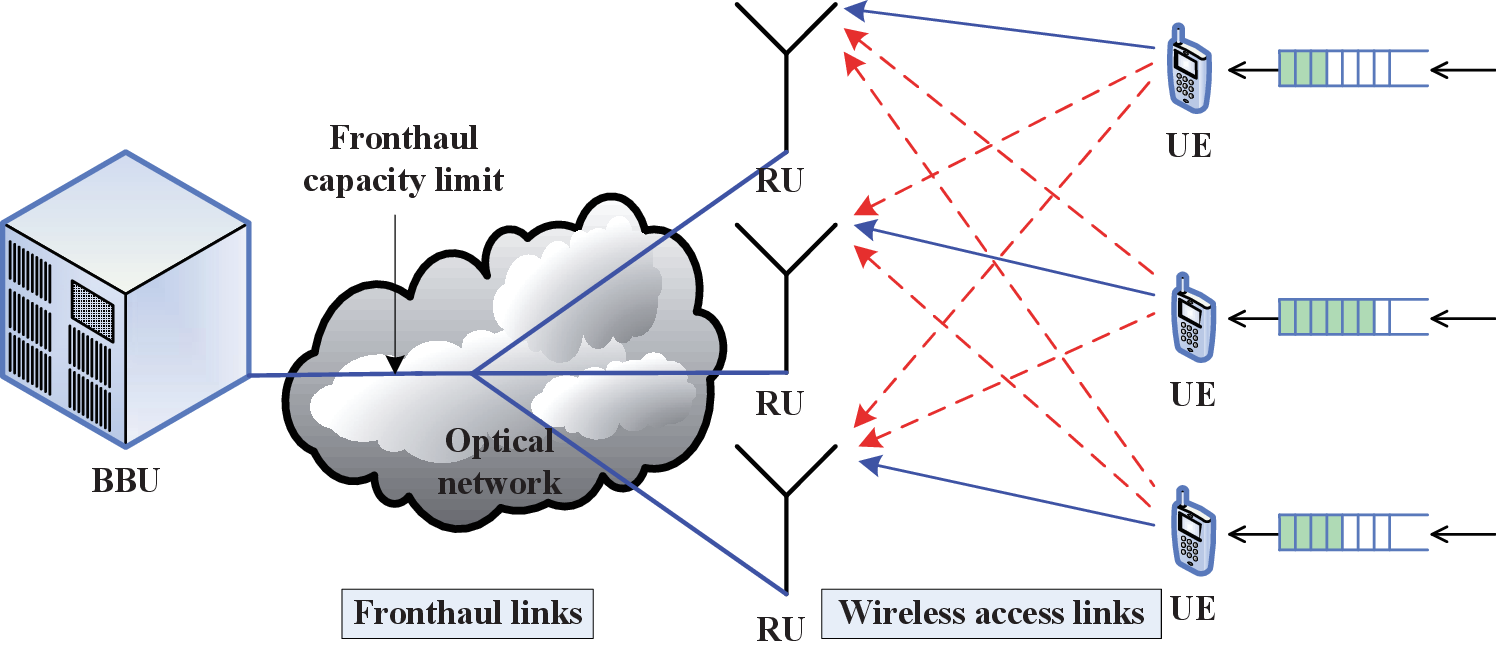}

\caption{C-RAN topology}
\label{fig:topology}
\end{figure}

We consider a C-RAN with $K$ cells, each of which has an RU with
a single antenna. In each cell, the data are transmitted from a single-antenna
user equipment (UE) to the RU via wireless access links and then to
the BBU via the fronthaul link over fiber/microwave, as shown in Fig.
\ref{fig:topology}.

The time is slotted and the duration of each time slot is $\tau$.
The BBU collects necessary information and makes the resource allocation
decisions periodically at the beginning of each time slot.

\subsection{Wireless Access Link Model}

The wireless access links are modeled as an interference channel.
In the uplink, the UEs transmit signals to their corresponding RUs
respectively, and in the meantime, cause interference to other RUs
in the network. The signals received by the RUs are
\begin{equation}
\mathbf{y}=\mathbf{H}\mathbf{x}+\mathbf{z},
\end{equation}
where $\mathbf{x}=(x_{1},x_{2},\cdots,x_{K}){}^{T}$ is a $K$-dimensional
vector of the transmitted signals, in which $x_{k}$ is transmitted
by the $k$-th UE\textcolor{red}{{} }with power $P$, $\mathbf{y}=(y_{1},y_{2},\cdots,y_{K}){}^{T}$
is a $K$-dimensional vector of the signals received by the RUs, in
which $y_{k}$ is the signal received by the RU in the $k$-th cell,
$\mathbf{H}=\left(H_{kj}\right)_{K\times K}$, in which $H_{kj}$
is the complex channel fading coefficient of the uplink transmission
from the $j$-th UE to the RU in the $k$-th cell, $\mathbf{z}=(z_{1},z_{2},\cdots,z_{K}){}^{T}$
and $z_{k}\sim\mathcal{{CN}}\left(0,N_{0}\right)$ is the white Gaussian
thermal noise with power $N_{0}$.

Define $\mathbf{H}(t)$ as the \emph{global CSI} for uplink access
links at the $t$-th slot. We have the following assumption on $\mathbf{H}(t)$.

\begin{assumption}[CSI Model]\label{Ass_csi}The CSI $\mathbf{H}(t)$
remains constant within a time slot and is i.i.d. over time slots.
$H_{kj}\left(t\right)$ is independent over the indices $k$ and $j$.%
\footnote{In C-RANs, the simultaneously transmitting UEs using the same resource
block are located in different cells. Thus, the distances between
the RUs and those between the UEs are always large enough to make
the channel fading coefficients independent.%
} $H_{kj}\left(t\right)$ is composed of two parts, i.e., $H_{kj}\left(t\right)=\sqrt{L_{kj}}\widetilde{H}_{kj}(t)$,
where $\widetilde{H}_{kj}(t)$ is the short-term fading coefficient
which follows a complex Gaussian distribution with mean 0 and unit
variance, and $L_{kj}$ is the corresponding large-scale path gain,
which is constant over the duration of the communication session.\hfill\IEEEQED\end{assumption}

\subsection{Fronthaul Link Model}

Denote $C_{k}(t)$ as the capacity allocated to the fronthaul link
between the RU in the $k$-th cell and the BBU at the $t$-th slot.
Let $\mathbf{C}(t)=\left(C_{1}(t),C_{2}(t),\cdots,C_{K}(t)\right)$
be the uplink fronthaul allocation. With limited-capacity fronthaul
links, the signals transmitted between the RUs and the BBU have to
be quantized. In the uplink, the RU in each cell underconverts its
received signal and sends the quantized signal to the BBU. Define
$\mathbf{\widehat{y}}=\left(\widehat{y}_{1},\widehat{y}_{2},\cdots,\widehat{y}_{K}\right){}^{T}$,
where $\widehat{y}_{k}$ is the quantized signal at the RU in the
$k$-th cell. The signals are assumed to be quantized for each fronthaul
link separately. The quantization leads to the distortion of signal,
which can be treated as the quantization noise, denoted as $\mathbf{n}=(n_{1},n_{2},\cdots,n_{K}){}^{T}$,
where $n_{k}$ is the quantization noise over the $k$-th fronthaul
link. The signals received by the BBU are expressed as
\begin{equation}
\mathbf{\widehat{y}}=\mathbf{y}+\mathbf{n}.
\end{equation}

The relationship between $y_{k}$ and $\hat{y}_{k}$ depends on the
fronthaul capacity $C_{k}$ according to the rate-distortion theory
\cite{info}, which is given by $I\left(y_{k}:\hat{y}_{k}\right)\leq C_{k}$,
where $I\left(y_{k}:\hat{y}_{k}\right)$ is the mutual information
between $y_{k}$ and $\hat{y}_{k}$. Let $\mathbf{N}(t)=\left(N_{1}(t),N_{2}(t),\cdots,N_{K}(t)\right)$,
where $N_{k}(t)$ is the power of the quantization noise $n_{k}$
at the $t$-th slot. The quantization noise power induced by the transmission
over the $k$-th uplink fronthaul link at the $t$-th slot is given
by \cite{WZcompression}
\begin{equation}
N_{k}\left(t\right)=\frac{P\sum_{j=1}^{K}\left\Vert H_{kj}\left(t\right)\right\Vert ^{2}+N_{0}}{2^{C_{k}\left(t\right)}-1},\label{eq:quantization}
\end{equation}
where $\left\Vert \bullet\right\Vert $ is the Euclidean norm.

\subsection{Throughput with Cloud Baseband Processing}

The BBU performs joint decoding for the received uplink signals, which
benefits the system performance by joint cloud processing of the signals
for different cells. The cloud baseband processing for uplink signals
at the BBU is introduced in the following assumption.

\begin{assumption}[Zero Forcing Joint Detection]\label{Ass_ZF}Assume
that ZF joint detection \cite{JD1,JD2} is adopted for the uplink
in the cloud baseband processing to eliminate the inter-cell interference.
The linear ZF receiver at the BBU can be represented by a matrix $\mathbf{S}(t)=\left(S_{kj}(t)\right)_{K\times K}$
at the $t$-th slot, where $\mathbf{S}(t)$ is the inverse%
\footnote{According to Assumption \ref{Ass_csi}, the elements of $\mathbf{H}(t)$
are independent. Thus, $\mathrm{rank}\left(\mathbf{H}(t)\right)=K,\forall t$
and the inverse of $\mathbf{H}(t)$ exists.%
} of the channel matrix $\mathbf{H}(t)$, i.e., $\mathbf{S}(t)=\mathbf{H}(t)^{-1}$.\hfill\IEEEQED\end{assumption}

\begin{figure}[t]
\centering

\includegraphics[width=5in]{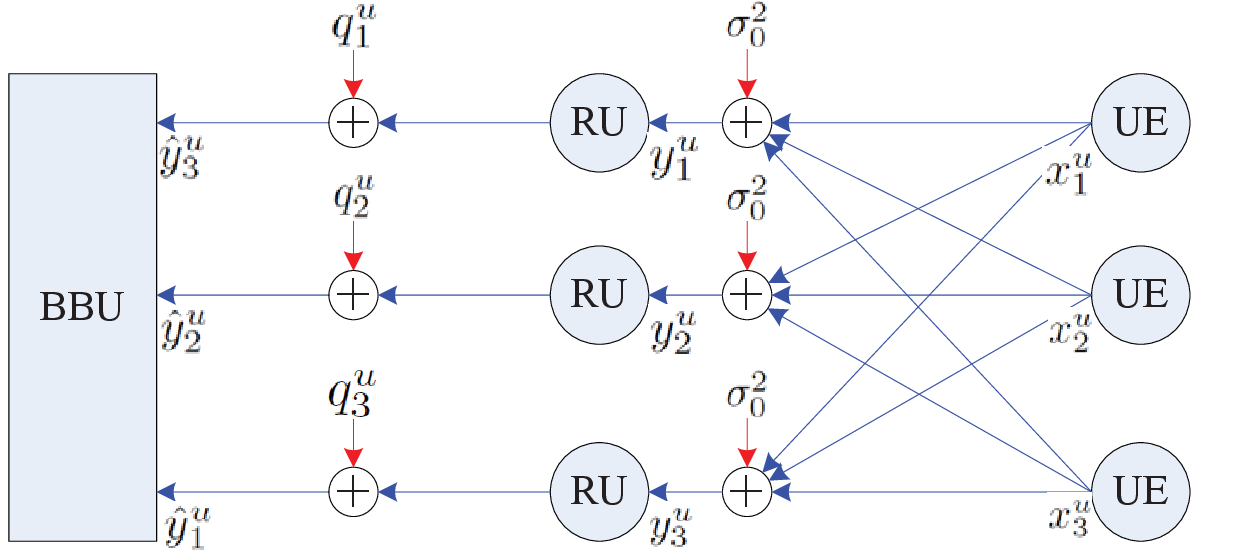}

\caption{Uplink transmission model of a C-RAN}
\label{fig:model}
\end{figure}

The uplink transmission model is described in Fig. \ref{fig:model}.
With the ZF joint detection at the BBU, the post-processing signal
is 
\begin{equation}
\mathbf{S}\mathbf{\hat{y}}=\mathbf{x}+\mathbf{S}(\mathbf{z}+\mathbf{q}).
\end{equation}

Considering both the thermal noise power $N_{0}$ and the quantization
noise power $\mathbf{N}(t)$, we obtain the uplink data rate for the
$i$-th UE as 
\begin{equation}
\begin{aligned}R_{k}\left(\mathbf{H}\left(t\right),\mathbf{C}\left(t\right)\right)= & \log_{2}\left(1+\frac{P}{\sum_{j=1}^{K}||S_{kj}(t)||^{2}\left(N_{0}+N_{j}\left(t\right)\right)}\right),\end{aligned}
\label{eq:rate}
\end{equation}
where $\mathbf{N}(t)$ is a function of $\mathbf{H}\left(t\right)$
and $\mathbf{C}\left(t\right)$, and $\mathbf{S}(t)$ is a function
of $\mathbf{H}\left(t\right)$. Note that there is an implicit coupling
among the $K$ uplink data flows in the sense that $R_{k}$ depends
not only on the fronthaul capacity allocation $C_{k}$ but also on
$C_{j},\forall j\neq k$.

\subsection{Queue Dynamics}

There is a bursty data source for each UE. Let $\mathbf{A}(t)=(A_{1}(t)\tau,\cdots,A_{N}(t)\tau)$
be the random arrivals (number of bits) from the application layers
at the end of the $t$-th time slot%
\footnote{We assume that the transmitters are causal so that the packets arrived
at the time slot are not observed when the control actions of this
time slot are performed.%
}. We have the following assumption on $\mathbf{A}(t)$. 

\begin{assumption}[Bursty Source Model]Assume that $A_{k}\left(t\right)$
is i.i.d. over slots according to a general distribution $\Pr[A_{k}]$.
The moment generating functions of $A_{k}$ exist with $\mathbb{E}[A_{k}]=\lambda_{k}$.
$A_{k}\left(t\right)$ is independent w.r.t. $k$. Furthermore, the
arrival rates $\boldsymbol{\lambda}=(\lambda_{1},\lambda_{2},\dots,\lambda_{K})$
lie within the stability region \cite{neely} of the system with the
given uplink resource.\hfill\IEEEQED\end{assumption}

Each UE has a data queue for the bursty traffic flows towards the
associated RU. Let $Q_{k}(t)\in[0,\infty)$ be the queue length (number
of bits) at the $k$-th UE at the beginning of the $t$-th slot. Let
$\mathbf{Q}(t)=(Q_{1}(t),\cdots,Q_{N}(t))\in\boldsymbol{\mathcal{Q}}\triangleq[0,\infty)^{K}$
be the \emph{global QSI}. The queue dynamics for the $k$-th UE can
be written as
\begin{equation}
Q_{k}(t+1)=\max\left\{ Q_{k}(t)-R_{k}(\mathbf{H}(t),\mathbf{C}(t))\tau,0\right\} +A_{k}(t)\tau.\label{eq:queue}
\end{equation}

\begin{remark}[Coupling Property of Uplink Queue Dynamics]In the
uplink, the $K$ queue dynamics are coupled together due to the ZF
processing in the BBU. Specifically, according to (\ref{eq:rate}),
the queue departure $R_{k}(\mathbf{H}(t),\mathbf{C}(t))$ for the
$i$-th UE depends on not only the allocated capacity $C_{k}(t)$
for the $k$-th fronthaul link, but also all the other elements of
$\mathbf{C}(t)$.\hfill\IEEEQED\end{remark}

\section{A Control Framework of Delay-Aware Uplink Fronthaul Allocation}

In this section, we formulate the delay-aware control framework of
uplink fronthaul allocation. We first define the control policy and
the optimization objective. We then formulate the design as a Markov
Decision Process (MDP) and derive the optimality conditions for solving
the problem.

\subsection{Fronthaul Allocation Policy}

For delay-sensitive applications, it is important to dynamically adapt
the fronthaul capacities $\mathbf{C}(t)$ based on the instantaneous
realizations of the CSI (captures the instantaneous transmission opportunities)
and the QSI (captures the urgency of $K$ data flows). Let $\boldsymbol{\chi}=(\mathbf{H},\mathbf{Q})$
denote the global system state. We define the stationary fronthaul
allocation policy below:

\begin{definition}[Stationary Fronthaul Allocation Policy] A stationary
control policy for the $k$-th UE $\Omega_{k}$ is a mapping from
the system state $\boldsymbol{\chi}$ to the fronthaul allocation
action of the $k$-th UE. Specifically, $\Omega_{k}(\boldsymbol{\chi})=C_{k}\geq0$.
Let $\boldsymbol{\Omega}=\{\Omega_{k}:\forall k\}$ denote the aggregation
of the control policies for all the $K$ UEs.\hfill \IEEEQED \end{definition}

The CSI $\mathbf{H}$ is i.i.d. over time slots based on the block
fading channel model in Assumption \ref{Ass_csi}. Furthermore, from
the queue evolution equation in (\ref{eq:queue}), $\mathbf{Q}(t+1)$
depends only on $\mathbf{Q}(t)$ and the data rate. Given a control
policy $\boldsymbol{\Omega}$, the data rate at the $t$-th slot depends
on $\mathbf{H}(t)$ and $\boldsymbol{\Omega}(\boldsymbol{\chi(t)})$.
Hence, the global system state $\boldsymbol{\chi}(t)$ is a controlled
Markov chain \cite{DPcontrol} with the following transition probability:
\begin{equation}
\begin{aligned}\Pr[\boldsymbol{\chi}(t+1)|\boldsymbol{\chi}(t),\boldsymbol{\Omega}(\boldsymbol{\chi}(t))]= & \Pr[\mathbf{H}(t+1)]\Pr[\mathbf{Q}(t+1)|\boldsymbol{\chi}(t),\boldsymbol{\Omega}(\boldsymbol{\chi}(t))]\end{aligned}
,
\end{equation}
where the queue transition probability is given by
\begin{equation}
\begin{aligned}\Pr[\mathbf{Q}(t+1)|\boldsymbol{\chi}(t),\boldsymbol{\Omega}(\boldsymbol{\chi}(t))]= & \begin{cases}
\prod_{k}\Pr\big[A_{k}\left(t\right)\big] & \text{if }Q_{k}\left(t+1\right)\text{is given by (\ref{eq:queue})},\forall k\\
0 & \text{otherwise},
\end{cases}\end{aligned}
\end{equation}
where the equality is due to the i.i.d. assumption of $\mathbf{H}(t)$
in Assumption \ref{Ass_csi}.

For technical reasons, we consider the \emph{admissible control policy}
defined below.

\begin{definition}[Admissible Control Policy]\label{Def_adm}A policy
$\boldsymbol{\Omega}$ is admissible if the following requirements
are satisfied:
\begin{itemize}
\item $\boldsymbol{\Omega}$ is a unichain policy, i.e., the controlled
Markov chain $\boldsymbol{\chi}\left(t\right)$ under $\boldsymbol{\Omega}$
has a single recurrent class (and possibly some transient states)
\cite{DPcontrol}.
\item The queueing system under $\boldsymbol{\Omega}$ is second-order stable
in the sense that $\lim_{t\rightarrow\infty}\mathbb{E}^{\boldsymbol{\Omega}}[\sum_{k=1}^{K}Q_{k}^{2}(t)]<\infty$,
where $\mathbb{E}^{\boldsymbol{\Omega}}$ means taking expectation
w.r.t. the probability measure induced by the control policy $\boldsymbol{\Omega}$.\hfill\IEEEQED
\end{itemize}
\end{definition}

\subsection{Problem Formulation}

As a result, under an admissible control policy $\boldsymbol{\Omega}$,
the average delay for the $k$-th data queue is given by
\begin{equation}
\overline{D}_{k}(\boldsymbol{\Omega})=\limsup_{T\rightarrow\infty}\frac{1}{T}\sum_{t=0}^{T-1}\mathbb{E}^{\boldsymbol{\Omega}}\left[\frac{Q_{k}\left(t\right)}{\lambda_{k}}\right],\quad\forall k.\label{eq:delaycost}
\end{equation}
Similarly, under an admissible control policy $\boldsymbol{\Omega}$,
the average fronthaul capacity for the $k$-th data queue is given
by

\begin{equation}
\overline{C}_{k}(\boldsymbol{\Omega})=\limsup_{T\rightarrow\infty}\frac{1}{T}\sum_{t=0}^{T-1}\mathbb{E}^{\boldsymbol{\Omega}}\left[C_{k}\left(t\right)\right],\quad\forall k.\label{eq:delaycost-1}
\end{equation}

We formulate the delay-aware fronthaul allocation problem for C-RANs
as follows:

\begin{problem}[Delay-Aware Fronthaul Allocation Problem]\label{Pro_MDP}The
delay-aware fronthaul allocation problem is formulated as
\begin{equation}
\begin{aligned}\underset{\boldsymbol{\Omega}}{\min}\quad & L(\boldsymbol{\Omega})=\sum_{k=1}^{K}\Big(\beta_{k}\overline{D}_{k}(\boldsymbol{\Omega})+\gamma_{k}\overline{C}_{k}(\boldsymbol{\Omega})\Big)\\
 & \qquad\;=\limsup_{T\rightarrow\infty}\frac{1}{T}\sum_{t=0}^{T-1}\mathbb{E}^{\boldsymbol{\Omega}}\left[c\left(\mathbf{Q}\left(t\right),\boldsymbol{\Omega}\left(\boldsymbol{\chi}\left(t\right)\right)\right)\right]
\end{aligned}
\end{equation}
where $c\left(\mathbf{Q},\mathbf{C}\right)=\sum_{k=1}^{K}\left(\beta_{k}\frac{Q_{k}}{\lambda_{k}}+\gamma_{k}C_{k}\right)$.
$\boldsymbol{\beta}=\{\beta_{k}>0:\forall k\}$ are the positive weights
for the delay cost and $\boldsymbol{\gamma}=\{\gamma_{k}>0:\forall k\}$
are the prices for the data transmission over fronthaul links. \hfill\IEEEQED\end{problem}

Note that Problem \ref{Pro_MDP} is an infinite horizon average cost
MDP, which is known as a very difficult problem.

\subsection{Optimality Conditions for Uplink Fronthaul Allocation}

Problem \ref{Pro_MDP} is an MDP and the \emph{Bellman equation} \cite{DPcontrol}
provides its optimality conditions. The Bellman equation\emph{ }involves
the entire system state $\boldsymbol{\chi}=(\mathbf{H},\mathbf{Q})$.
Exploiting the i.i.d. property of $\mathbf{H}(t)$ according to Assumption
\ref{Ass_csi}, we obtain the \emph{equivalent Bellman equation} in
the following theorem.

\begin{theorem}[Sufficient Conditions for Optimality]\label{The_opt}For
any given weights $\boldsymbol{\beta}$, assume there exists a $\left(\theta^{*},\left\{ V^{*}(\mathbf{Q})\right\} \right)$
that solves the following \emph{equivalent Bellman equation}:
\begin{equation}
\begin{aligned} & \theta^{*}\tau+V^{*}(\mathbf{Q})=\mathbb{E}\bigg[\min_{\boldsymbol{\Omega}(\boldsymbol{\chi})}\Big[c\big(\mathbf{Q},\boldsymbol{\Omega}\big(\boldsymbol{\chi}\big)\big)\tau+\sum_{\mathbf{Q}'}\Pr\big[\mathbf{Q}'\big|\boldsymbol{\chi},\boldsymbol{\Omega}\big(\boldsymbol{\chi}\big)\big]V^{*}(\mathbf{Q}')\Big]\bigg|\mathbf{Q}\bigg],\:\forall\mathbf{Q}\in\boldsymbol{\mathcal{Q}},\end{aligned}
\label{eq:bellman1}
\end{equation}
Furthermore, for all admissible control policies $\boldsymbol{\Omega}$,
$V^{\ast}$ satisfies the following \emph{transversality condition}:
\begin{equation}
\lim_{T\rightarrow\infty}\frac{1}{T}\mathbb{E}^{\boldsymbol{\Omega}}\left[V^{\ast}\left(\mathbf{Q}\left(T\right)\right)\right]=0.\label{eq:trans1}
\end{equation}
Then $\theta^{*}$\textcolor{black}{{} }is the optimal average cost,
and $V^{\ast}\left(\mathbf{Q}\right)$ is the \emph{priority function}
of the $K$ data flows. If there exists an admissible stationary policy
$\boldsymbol{\Omega}^{\ast}\left(\boldsymbol{\chi}\right)=\mathbf{C}^{*}$
where $\mathbf{C}^{*}$ attains the minimum of the R.H.S. of (\ref{eq:bellman1})
for all $\mathbf{Q}\in\boldsymbol{\mathcal{Q}}$, then $\boldsymbol{\Omega}^{*}$
is the optimal control policy for Problem \ref{Pro_MDP}.\end{theorem}

\begin{IEEEproof}Please refer to Appendix A.\end{IEEEproof} 

\begin{remark} [Interpretation of Theorem \ref{The_opt}]The equivalent
Bellman equation in (\ref{eq:bellman1}) is defined on the QSI $\mathbf{Q}$
only. Nevertheless, the optimal control policy $\boldsymbol{\Omega}^{\ast}$
obtained by solving (\ref{eq:bellman1}) is still adaptive to the
entire system state $\boldsymbol{\chi}$. At each stage, when the
queue length is $\mathbf{Q}(t)$, the optimal action has to strike
a balance between the current cost $c\big(\mathbf{Q},\boldsymbol{\Omega}\big(\boldsymbol{\chi}\big)\big)$
and the future cost $\sum_{\mathbf{Q}'}\Pr\big[\mathbf{Q}'\big|\boldsymbol{\chi},\boldsymbol{\Omega}\big(\boldsymbol{\chi}\big)\big]V^{*}(\mathbf{Q}')$
because the action taken will affect the future evolution of $\mathbf{Q}(t+1)$.\hfill\IEEEQED \end{remark}

\section{Low-Complexity Fronthaul Allocation}

One key obstacle in deriving the optimal fronthaul policy $\boldsymbol{\Omega}^{\ast}$
is to obtain the priority function $V^{\ast}(\mathbf{Q})$ of the
Bellman equation in (\ref{eq:bellman1}). Conventional brute force
value iteration or policy iteration algorithms can only give numerical
solutions and have exponential complexity in $K$, which is highly
undesirable. In this section, we shall exploit the characteristics
of the topology of C-RANs. Specifically, we define $\delta=\max\left\{ L_{kj}:\forall k\neq j\right\} $
be the worst-case path gain among all the cross links, which is usually
weaker than the home cell path gain due to the C-RAN network architecture.
We adopt perturbation theory w.r.t. $\delta$ to obtain a closed-form
approximation of the priority function $V^{\ast}(\mathbf{Q})$ and
derive the associated error bound. Based on that, we obtain a low
complexity delay-aware fronthaul allocation algorithm.

\subsection{Calculus Approach for Solving the Bellman Equation}

We adopt a calculus approach to obtain a closed-form approximate priority
function. We first have the following theorem for solving the Bellman
equation in (\ref{eq:bellman1}). 

\begin{theorem}[Calculus Approach for Solving (\ref{eq:bellman1})]\label{The_HJB1}Assume
there exist $c^{\infty}$ and $J\left(\mathbf{Q};\delta\right)$ of
class $\mathcal{C}^{2}(\mathbb{R}_{+}^{K})$ that satisfy
\begin{itemize}
\item the following partial differential equation (PDE):
\begin{equation}
\begin{aligned} & \mathbb{E}\Bigg[\min_{\boldsymbol{\Omega}\left(\boldsymbol{\chi}\right)}\bigg[\sum_{k=1}^{K}\left(\beta_{k}\frac{Q_{k}}{\lambda_{k}}+\gamma_{k}C_{k}\right)-c^{\infty}+\sum_{k=1}^{K}\bigg(\frac{\partial J\left(\mathbf{Q};\delta\right)}{\partial Q_{k}}\left(\lambda_{k}-R_{k}\big(\mathbf{H},\mathbf{C}\big)\right)\bigg)\bigg]\Bigg|\mathbf{Q}\Bigg]=0,\\
 & \hspace{5cm}\hspace{5cm}\forall\mathbf{Q}\in\mathbb{R}_{+}^{K}
\end{aligned}
\label{eq:bellman3}
\end{equation}
with boundary condition $J\left(\mathbf{0};\delta\right)=0$;
\item For all $k$, $\frac{\partial J\left(\mathbf{Q};\delta\right)}{\partial Q_{k}}$
is an increasing function of all $Q_{k}$;
\item $J\left(\mathbf{Q};\delta\right)=\mathcal{O}\left(\|\mathbf{Q}\|^{2}\right)$.
\end{itemize}
Then, we have
\begin{equation}
\theta^{\ast}=c^{\infty}+o(1),V^{\ast}\left(\mathbf{Q}\right)=J\left(\mathbf{Q};\delta\right)+o(1),\forall\mathbf{Q}\in\boldsymbol{\mathcal{Q}},\label{eq:15resu}
\end{equation}
where the error term $o(1)$ asymptotically goes to zero for sufficiently
small $\tau$.\end{theorem}

\begin{IEEEproof}Please refer to Appendix B. \end{IEEEproof}

Theorem \ref{The_HJB1} suggests that if we can solve for the PDE
in (\ref{eq:bellman3}), then the solution $\left(J\left(\mathbf{Q};\delta\right),c^{\infty}\right)$
is only $o(1)$ away from the solution of the Bellman equation $(V^{*}(\mathbf{Q}),\theta^{*})$.

\subsection{Closed-Form Approximate Priority Function via Perturbation Analysis}

The queues of the $K$ uplink data flows are coupled due to the coupling
of $R_{k}$ in (\ref{eq:rate}). The following lemma establishes the
intensity of the queue coupling.

\begin{lemma}[Intensity of the Uplink Queue Coupling]\label{Lem_weak}The
coupling intensity of uplink data queues induced by $R_{k}$ in (\ref{eq:rate})
is given by $||S_{kj}(t)||^{2}=\mathcal{O}\left(\delta\right),\forall k\neq j$.\end{lemma}

\begin{IEEEproof}Please refer to Appendix C.\end{IEEEproof}

As a result, the solution of (\ref{eq:bellman3}) depends on the worst-case
cross link interference path gain $\delta$ and, hence, the $K$-dimensional
PDE in (\ref{eq:bellman3}) can be regarded as a perturbation of a
\emph{base system}, as defined below.

\begin{definition}[Base System]\label{Def_base}A base system is
characterized by the PDE in (\ref{eq:bellman3}) with $\delta=0$.\hfill \IEEEQED \end{definition}

According to Lemma \ref{Lem_weak}, we have $||S_{kj}(t)||^{2}=0,\forall k\neq j$
in the base system. We first study the base system and use $J(\mathbf{Q};0)$
to obtain a closed-form approximation of $J(\mathbf{Q};\delta)$.

We have the following lemma summarizing the priority function $J(\mathbf{Q};0)$
of the base system.

\begin{lemma} [Decomposable Structure of $J(\mathbf{Q};0)$]\label{Lem_base}The
solution $J(\mathbf{Q};0)$ for the base system has the following
decomposable structure:
\begin{equation}
J\left(\mathbf{Q};0\right)=\sum_{k=1}^{K}J_{k}\left(Q_{k}\right),\label{eq:linearA}
\end{equation}
where $J_{k}\left(Q_{k}\right)$ is the \emph{per-flow priority function}
for the $k$-th data flow given by 
\begin{equation}
\left\{ \begin{aligned}Q_{k}(\nu)= & \frac{\lambda_{k}}{\beta_{k}}\left(\frac{\nu e^{a_{k}}}{\ln2}E_{1}\left(\frac{a_{k}\nu}{\nu-\gamma_{k}}\right)-\lambda_{k}\nu\right.\left.-\frac{\gamma_{k}}{\ln2}E_{1}\left(\frac{a_{k}\gamma_{k}}{\nu-\gamma_{k}}\right)+c_{k}^{\infty}\right)\\
J_{k}(\nu)= & \frac{\lambda_{k}}{2\beta_{k}\ln2}\left(\gamma_{k}\left(\gamma_{k}-\nu\right)e^{\frac{a_{k}\gamma_{k}}{\gamma_{k}-\nu}}+e^{a_{k}}\nu^{2}E_{1}\left(\frac{a_{k}\nu}{\nu-\gamma_{k}}\right)\right.\\
 & \left.+\left(a_{k}-\lambda_{k}\right)E_{1}\left(\frac{a_{k}\gamma_{k}}{\nu-\gamma_{k}}\right)-\lambda_{k}\nu^{2}\ln2\right)+b_{k},
\end{aligned}
\right.\label{eq:perflow}
\end{equation}
where $a_{k}\triangleq\frac{N_{0}}{PL_{kk}}$; $c_{k}^{\infty}=\frac{\gamma_{k}}{\ln2}E_{1}\left(\frac{a_{k}\gamma_{k}}{d_{k}-\gamma_{k}}\right)$,
where $d_{k}$ satisfies $\frac{e^{a_{k}}}{\ln2}E_{1}\left(\frac{a_{k}d_{k}}{d_{k}-\gamma_{k}}\right)=\lambda_{k}$;
$E_{1}(z)\triangleq\int_{1}^{\infty}\frac{e^{-tz}}{t}\mathrm{d}t=\int_{z}^{\infty}\frac{e^{-t}}{t}\mathrm{d}t$;
$b_{k}$ is chosen to satisfy%
\footnote{To find $b_{k}$, firstly solve $Q_{k}(\nu)=0$ using one-dimensional
search techniques (e.g., bisection method). Then $b_{k}$ is chosen
such that $J_{k}(\nu)=0$.%
} the boundary condition $J_{k}(0)=0$.\end{lemma}

\begin{IEEEproof}Please refer to Appendix D.\end{IEEEproof}

Note that when $\delta=0$, we have $L_{kj}=0$ for all $k\neq j$
and, hence, there is no coupling between the UE-RU pairs. As a result,
the $K$ data queues are totally decoupled and the system is equivalent
to a decoupled system with $K$ independent queues. That is why the
priority function $J\left(\mathbf{Q};0\right)$ in the base system
has the decomposable structure in Lemma \ref{Lem_base}.

When $\delta>0$, $J(\mathbf{Q};\delta)$ can be considered as a perturbation
of the solution of the base system $J(\mathbf{Q};0)$. Using perturbation
analysis on the PDE (\ref{eq:bellman3}), we establish the following
theorem on the approximation of $J(\mathbf{Q};\delta)$:

\begin{theorem} [First Order Perturbation of $J(\mathbf{Q}; \delta)$]\label{The_first}$J(\mathbf{Q};\delta)$
is given by
\begin{equation}
\begin{aligned}J\left(\mathbf{Q};\delta\right)= & J\left(\mathbf{Q};0\right)+\sum_{k=1}^{K}\left(\sum_{j=1,j\neq k}^{K}L_{kj}\left(\Phi_{k}Q_{k}^{2}\sum\limits _{l=1,l\neq k}^{K}\frac{N_{0}}{L_{ll}}\right.\right.\left.\left.+\frac{N_{0}}{L_{jj}}\sum\limits _{i=1,i\neq k,j}^{K}\Phi_{i}Q_{i}^{2}\right)+o\left(Q_{k}^{2}\right)\right)+\mathcal{O}\left(\frac{1}{\delta^{2}}\right),\end{aligned}
\label{eq:firstorder}
\end{equation}
where $\Phi_{k}=\frac{\beta_{k}}{\lambda_{k}}\left(1-\frac{a_{k}e^{a_{k}}E_{1}\left(a_{k}\right)}{N_{0}}\right)/\left(e^{a_{k}}E_{1}\left(a_{k}\right)-\lambda_{k}\ln2\right)$.\end{theorem}

\begin{IEEEproof}Please refer to Appendix E.\end{IEEEproof}

The priority function $V(\mathbf{Q})$ is decomposed into the following
three terms: 1) the base term $\sum_{k}J_{k}(Q_{k})$ obtained by
solving a base system without coupling, 2) the perturbation term accounting
for the first order coupling due to the joint processing in the BBU,
and 3) the residual error term which goes to zero in the order of
$\mathcal{O}(1/\delta^{2})$. As a result, we adopt the following
closed-form approximation of $V(\mathbf{Q})$:
\begin{equation}
\begin{aligned}\widetilde{V}(\mathbf{Q})= & \sum_{k=1}^{K}J_{k}(Q_{k})+\sum_{k=1}^{K}\left(\sum_{j=1,j\neq k}^{K}L_{kj}\left(\Phi_{k}Q_{k}^{2}\sum\limits _{l=1,l\neq k}^{K}\frac{N_{0}}{L_{ll}}\right.\right.\left.\left.+\frac{N_{0}}{L_{jj}}\sum\limits _{i=1,i\neq k,j}^{K}\Phi_{i}Q_{i}^{2}\right)\right).\end{aligned}
\label{eq:finalapprox}
\end{equation}

\subsection{Fronthaul Allocation Algorithm}

In this section, we use the closed-form approximate priority function
in (\ref{eq:finalapprox}) to capture the urgency information of the
$K$ data flows and obtain a low complexity delay-aware fronthaul
allocation algorithm. Using the approximate priority function in (\ref{eq:finalapprox}),
the per-stage control problem (for each state realization $\boldsymbol{\chi}$)
is given by%
\footnote{Note that $J_{k}'\left(Q_{k}\right)=\left(\frac{\mathrm{d}J_{k}\left(\nu\right)}{\mathrm{d}\nu}\Big/\frac{\mathrm{d}Q_{k}\left(\nu\right)}{\mathrm{d}\nu}\right)\Big|_{\nu=\nu\left(Q_{k}\right)}=\nu\left(Q_{k}\right)$,
where $\nu\left(Q_{k}\right)$ satisfies $Q_{k}\left(\nu\left(Q_{k}\right)\right)=Q_{k}$.%
}
\begin{equation}
\max_{\mathbf{C}}\ \sum_{k=1}^{K}\left(\frac{\partial\widetilde{V}\left(\mathbf{Q}\right)}{\partial Q_{k}}R_{k}\left(\mathbf{H},\mathbf{C}\right)-\gamma_{k}C_{k}\right),\label{eq:utility}
\end{equation}
where $\frac{\partial\widetilde{V}\left(\mathbf{Q}\right)}{\partial Q_{k}}$
can be calculated from (\ref{eq:finalapprox}), which is given by
\begin{equation}
\begin{aligned}\frac{\partial\widetilde{V}\left(\mathbf{Q}\right)}{\partial Q_{k}}= & J_{k}'\left(Q_{k}\right)+2\Phi_{k}\left(\sum_{j=1,j\neq k}^{K}L_{kj}\sum\limits _{l=1,l\neq k}^{K}\frac{N_{0}}{L_{ll}}\right.\left.+\sum_{i=1,i\neq k}^{K}\sum_{j=1,j\neq i,k}^{K}L_{ij}\frac{N_{0}}{L_{jj}}\right)Q_{k}.\end{aligned}
\end{equation}

The per-stage problem in (\ref{eq:utility}) is similar to the weighted
sum-rate (WSR) optimization \cite{weiyu}, which can be considered
as a special case of network utility maximization. However, unlike
conventional WSR problems, where the weights are static, the weights
here in (\ref{eq:utility}) are dynamic and are determined by the
QSI via the priority function $\frac{\partial\widetilde{V}(\mathbf{Q})}{\partial Q_{k}}$.
As such, the role of the QSI is to dynamically adjust the weight (priority)
of the individual flows, whereas the role of the CSI is to adjust
the priority of the flow based on the transmission opportunity in
the rate function $R_{k}(\mathbf{H},\mathbf{C})$.

One approach to solve the WSR problem is solving the local optimization
problem for each flow iteratively \cite{weiyu}. In each local optimization
problem for the $k$-th flow, the total WSR objective is maximized,
assuming that the capacities of other links $C_{j},\forall j\neq k$
do not change. The local optimization problem is formulated as
\begin{equation}
\max_{C_{k}}\ \sum_{k=1}^{K}\left(\frac{\partial\widetilde{V}\left(\mathbf{Q}\right)}{\partial Q_{k}}R_{k}\left(\mathbf{H},\mathbf{C}\right)-\gamma_{k}C_{k}\right).\label{eq:local}
\end{equation}
The above local optimization problem is still difficult to solve directly.
An alternative method is simplifying the effect of $C_{k}$ on the
other links as a linear function%
\footnote{We will show later that this simplification does not affect the convergence
property.%
} \cite{Huang}. Define $\pi_{ik}$ as the marginal increase in the
utility of the $i$-th flow per unit increase in $C_{k}$, i.e., 
\begin{equation}
\begin{aligned}\pi_{ik}=\frac{\frac{\partial\widetilde{V}\left(\mathbf{Q}\right)}{\partial Q_{i}}P\left\Vert S_{ik}\right\Vert ^{2}Y_{k}\frac{2^{C_{k}}}{\left(2^{C_{k}}-1\right)^{2}}}{\left(P+I_{ik}+\left\Vert S_{ik}\right\Vert ^{2}\left(N_{0}+\frac{Y_{k}}{2^{C_{k}}-1}\right)\right)\left(I_{ik}+\left\Vert S_{ik}\right\Vert ^{2}\left(N_{0}+\frac{Y_{k}}{2^{C_{k}}-1}\right)\right)},\end{aligned}
\label{eq:pi}
\end{equation}
where $I_{ik}=\sum_{j=1,j\neq k}^{K}\left\Vert S_{ij}\right\Vert ^{2}\left(N_{0}+\frac{Y_{j}}{2^{C_{j}}-1}\right)$
and $Y_{j}=P\sum_{l=1}^{K}\left\Vert H_{jl}\right\Vert ^{2}+N_{0}$. 

Adopting the linear simplification $\pi_{ik}C_{k}$ for the effect
of $C_{k}$ on the $i$-th flow in the per-stage local optimization
problem (\ref{eq:local}), we have the Karush-Kuhn-Tucker (KKT) condition
as
\begin{equation}
\begin{aligned} & \frac{\frac{\partial\widetilde{V}\left(\mathbf{Q}\right)}{\partial Q_{k}}\frac{P\left\Vert S_{kk}\right\Vert ^{2}Y_{k}2^{C_{k}}}{\left(2^{C_{k}}-1\right)^{2}}}{\left(I_{kk}+\left\Vert S_{kk}\right\Vert ^{2}\left(N_{0}+\frac{Y_{k}}{2^{C_{k}}-1}\right)\right)\left(P+I_{kk}+\left\Vert S_{kk}\right\Vert ^{2}\left(N_{0}+\frac{Y_{k}}{2^{C_{k}}-1}\right)\right)}+\sum_{i=1,i\neq k}^{K}\pi_{ik}=\gamma.\end{aligned}
\label{eq:local2}
\end{equation}
By solving (\ref{eq:local2}), we obtain the optimal fronthaul capacity
$C_{k}$ for the local optimization problem as
\begin{equation}
C_{k}=\log_{2}\frac{\eta_{k}+\zeta_{k}+\sqrt{\eta_{k}^{2}+2\eta_{k}\zeta_{k}+P^{2}\left\Vert S_{kk}\right\Vert ^{4}Y_{k}^{2}}}{2\left(P+I_{kk}+\left\Vert S_{kk}\right\Vert ^{2}N_{0}\right)\left(I_{kk}+\left\Vert S_{kk}\right\Vert ^{2}N_{0}\right)},\label{eq:local3}
\end{equation}
where $\zeta_{k}=2I_{kk}^{2}+2I_{kk}(P+2\left\Vert S_{kk}\right\Vert ^{2}N_{0}-\left\Vert S_{kk}\right\Vert ^{2}Y_{k})+\left\Vert S_{kk}\right\Vert ^{2}(2\left\Vert S_{kk}\right\Vert ^{2}N_{0}^{2}-PY_{k}+2PN_{0}-2\left\Vert S_{kk}\right\Vert ^{2}N_{0}Y_{k})$
and $\eta_{k}=\frac{\partial\widetilde{V}\left(\mathbf{Q}\right)}{\partial Q_{k}}\frac{P\left\Vert S_{kk}\right\Vert ^{2}Y_{k}}{\gamma-\sum_{i=1,,i\neq k}^{K}\pi_{ik}}$.

Based on the above analysis, we propose a low-complexity fronthaul
allocation algorithm launched at the beginning of each slot, which
is described using pseudo codes as Algorithm 1. We denote $\mathbf{C}^{(n)}=(C_{1}^{(n)},C_{2}^{(n)},\cdots,C_{K}^{(n)})$
as the allocated fronthaul capacities in the $n$-th iteration.

\begin{algorithm}[tbh]
\caption{Delay-Aware Fronthaul Allocation}

\begin{algorithmic}[1]
\STATE Initialize $n=0$ and $C^{(0)}_k=0,\forall k$
\REPEAT
\FOR{all user $k$}
\STATE Calculate $C^{(n+1)}_k$ based on $\mathbf{C}^{(n)}$ according to (\ref{eq:local3})
\ENDFOR
\STATE $n=n+1$
\UNTIL{The difference between $\mathbf{C}^{(n)}$ and $\mathbf{C}^{(n+1)}$ is below a given threshold}
\end{algorithmic}
\end{algorithm}

Although the per-stage problem (\ref{eq:utility}) is not convex in
general, the following lemma states that it is a convex problem for
sufficiently small $\delta$.

\begin{lemma}[Asymptotic Convexity]\label{The:convex}When $\delta$
is sufficiently small, the objective in (\ref{eq:utility}) is a concave
function of $\mathbf{C}$, and the problem (\ref{eq:utility}) is
a convex problem.\end{lemma}

\begin{IEEEproof}Please refer to Appendix F.\end{IEEEproof}

According to Lemma \ref{The:convex}, we provide the convergence property
and asymptotic optimality of Algorithm 1 in the following theorem: 

\begin{theorem}[Asymptotic Optimality]\label{The:convergence}When
$\delta$ is sufficiently small, starting from any feasible initial
point $\mathbf{C}^{(0)}$, Algorithm 1 converges to the optimal solution
of the original Problem \ref{Pro_MDP}.\end{theorem}

\begin{IEEEproof}Please refer to Appendix G.\end{IEEEproof}

\section{Simulation}

In this section, we evaluate the performance of the proposed low-complexity
delay-aware fronthaul allocation algorithm for C-RANs. For performance
comparison, we adopt the following two baseline schemes.
\begin{itemize}
\item \textbf{Baseline 1 {[}Throughput-Optimal Fronthaul Allocation{]}:}
The throughput-optimal fronthaul allocation algorithm determines the
fronthaul capacities for maximizing the total data rate without considering
the queueing information, which is similar to that in \cite{WZcompression}
but with ZF processing.
\item \textbf{Baseline 2 {[}Queue-Weighted Fronthaul Allocation{]}:} The
queue-weighted fronthaul allocation algorithm exploits both CSI and
QSI for queue stability by Lyapunov drift \cite{neely} and solves
the per-stage problem ($\ref{eq:utility}$) replacing $\frac{\partial\widetilde{V}(\mathbf{Q})}{\partial Q_{k}}$
with $Q_{k}$ \cite{MLWDF}.
\end{itemize}

In the simulation, the performance of the proposed fronthaul allocation
algorithm is evaluated in a C-RAN cluster with seven cells. A single
channel is considered, and one user over the channel is located randomly
in each cell, with radius 500m. Poisson data arrival is considered,
with an average arrival rate $\lambda_{k}$ for the $k$-th UE, which
is uniformly distributed between $[0,2\overline{\lambda}]$ with mean
$\overline{\lambda}$. The path gain is calculated as $L_{kj}=15.3+37.6\log_{10}d_{kj}$,
with the fading coefficient distributed as $\mathcal{CN}(0,1)$. The
average transmit power is 23dBm and the noise power spectrum density
is -174dBm/Hz. The system bandwidth is 10MHz and the duration of the
decision slot is 10ms. The weights $\gamma_{k}$ are the same and
$\beta_{k}=1$ for all $k$. For comparison, the delay performances
of different schemes are evaluated with the same total fronthaul capacity
by adjusting $\gamma_{k}$. For obtaining the average performance,
we consider 20 random topologies, each of which has 100 time slots.

\begin{figure}[t]
\centering\includegraphics[width=4.5in]{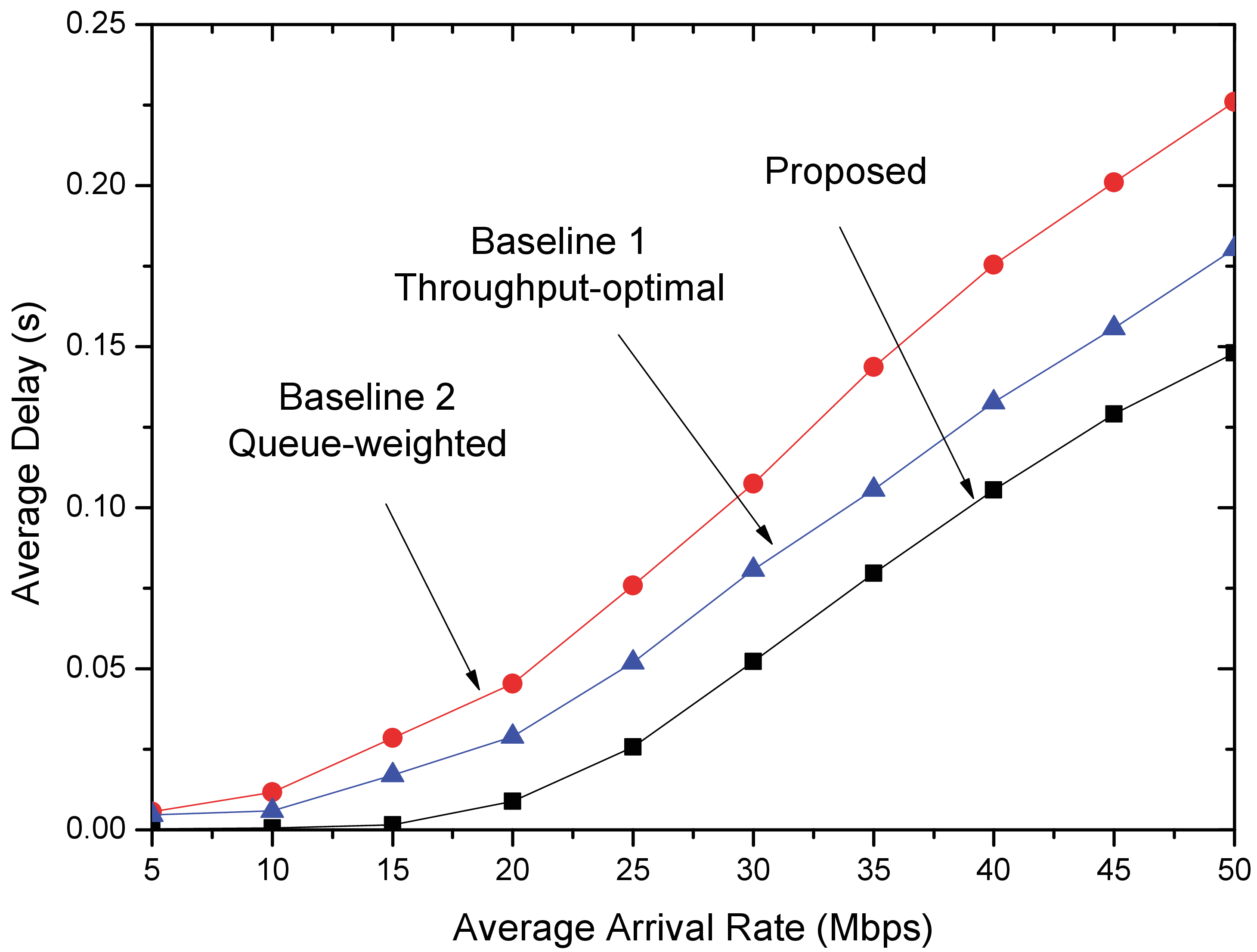}

\caption{Performance comparison with different average arrival rates when $\overline{C}=350\mathrm{Mbps}$ }
\label{fig:sim1}

\end{figure}

Fig. $\ref{fig:sim1}$ shows the average delay versus the average
arrival rate when the total fronthaul capacity is 350Mbps. For all
algorithms, the average delay increases when the average traffic load
increases. It can be observed that the proposed fronthaul allocation
algorithm outperforms both baselines, which verifies the accuracy
of the priority function approximation in the proposed algorithm.

\begin{figure}[t]
\centering\includegraphics[width=4.5in]{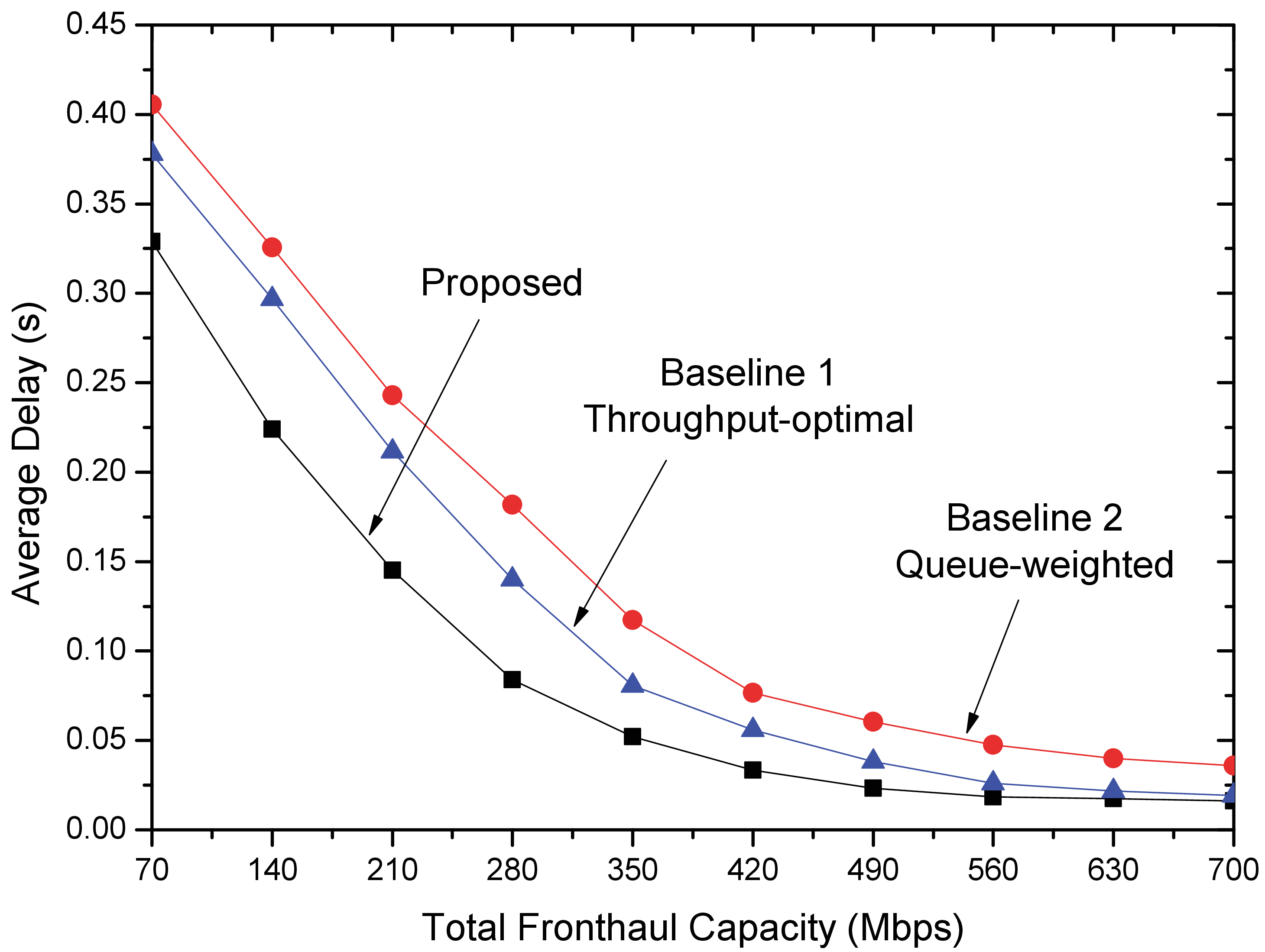}

\caption{Performance comparison with different total fronthaul capacity when
$\overline{\lambda}=30\mathrm{Mbps}$}
\label{fig:sim2}

\end{figure}

Fig. \ref{fig:sim2} shows the average delay versus the total fronthaul
capacity when the average arrival rate is 30Mbps. The proposed fronthaul
allocation algorithm also achieves better performance than the baseline
schemes. When the total fronthaul capacity is small, the average delay
decreases significantly with the increase of the total fronthaul capacity.
In contrast, when the total fronthaul capacity is large, the change
in the average delay is relatively small with adjustment of the total
fronthaul capacity.

Table \ref{tab:time} illustrates a comparison of the MATLAB computational
time of the proposed solution, the baselines and the brute-force value
iteration algorithm \cite{DPcontrol} in one time slot. From the results,
our proposed algorithm has much less complexity than the brute-force
value iteration algorithm. The computational time of our proposed
algorithm is close to those of Baselines 1 \& 2, and the difference
is due to the computation of the approximate priority function. Therefore,
our proposed algorithm achieves significant performance gain compared
to the baselines, with small computational complexity cost.

\begin{table}[t]
\caption{Comparison of the MATLAB computational time}
\label{tab:time}

\begin{center} 
\begin{tabular}{|l|l|}
\hline 
Algorithm & Time   \\
\hline
Baselines 1 \& 2 & 0.006s\\ 
Proposed Algorithm &  0.043s\\	 
Brute-Force Value Iteration &   $>10^5$s \\ 
\hline 
\end{tabular} 
\end{center}
\end{table}

\section{Conclusions}

In this paper, we propose a low-complexity delay-aware fronthaul allocation
algorithm for the uplink in C-RANs. The delay-aware fronthaul allocation
problem is formulated as an infinite horizon average cost Markov decision
process. To deal with the curse of dimensionality, we exploit the
specific problem structure that the cross link path gain is usually
weaker than the home cell path gain. Utilizing the\emph{ }perturbation
analysis\emph{ }technique, we obtain a closed-form approximate priority
function and the associated error bound. Based on the closed-form
approximate priority function, we propose a low-complexity delay-aware
fronthaul allocation algorithm, solving the per-stage optimization
problem. The proposed solution is further shown to be asymptotically
optimal for sufficiently small cross link path gains. The simulation
results verify the accuracy of the priority function approximation
and show that the proposed fronthaul allocation algorithm outperforms
the baselines.

\appendices{}

\section{Proof of Theorem \ref{The_opt}}

Following \emph{Prop. 4.6.1} of \cite{DPcontrol}, the sufficient
conditions for the optimality of \emph{Problem 1} are that $\left(\theta^{\ast},\{V^{\ast}\left(\mathbf{Q}\right)\}\right)$
solves the following Bellman equation:
\begin{equation}
\begin{aligned}\theta^{\ast}\tau+V^{\ast}\left(\boldsymbol{\chi}\right)= & \min_{\boldsymbol{\Omega}(\boldsymbol{\chi})}\Big[c\big(\mathbf{Q},\boldsymbol{\Omega}\big(\boldsymbol{\chi}\big)\big)\tau+\sum_{\boldsymbol{\chi}'}\Pr\big[\boldsymbol{\chi}'\big|\boldsymbol{\chi},\boldsymbol{\Omega}\big(\boldsymbol{\chi}\big)\big]V^{\ast}\left(\boldsymbol{\chi}'\right)\Big]\\
= & \min_{\boldsymbol{\Omega}(\boldsymbol{\chi})}\Big[c\big(\mathbf{Q},\boldsymbol{\Omega}\big(\boldsymbol{\chi}\big)\big)\tau+\sum_{\mathbf{Q}'}\sum_{\mathbf{H}'}\Pr\big[\mathbf{Q}'\big|\boldsymbol{\chi},\boldsymbol{\Omega}\big(\boldsymbol{\chi}\big)\big]\Pr\big[\mathbf{H}'\big]V^{\ast}\left(\boldsymbol{\chi}'\right)\Big]\Big]
\end{aligned}
\label{eq:app1bellman1}
\end{equation}
and $V^{\ast}$ satisfies the condition in (\ref{eq:trans1}) for
all admissible policies $\boldsymbol{\Omega}$. Then $\theta^{*}=\underset{\boldsymbol{\Omega}(\boldsymbol{\chi})}{\min}\, L\left(\boldsymbol{\Omega}\big(\boldsymbol{\chi}\big)\right)$. 

Taking expectation w.r.t. $\mathbf{H}$ on both sides of (\ref{eq:app1bellman1})
and denoting $V^{\ast}\left(\mathbf{Q}\right)=\mathbb{E}\big[V^{\ast}\left(\boldsymbol{\chi}\right)\big|\mathbf{Q}\big]$,
we obtain the equivalent Bellman equation in (\ref{eq:bellman1})
in Theorem \ref{The_opt}.

\section{Proof of Theorem \ref{The_HJB1}}

In the proof, we shall first establish the relationship between the
equivalent Bellman equation in (\ref{eq:bellman1}) in Theorem \ref{The_HJB1}
and the approximate Bellman equation in (\ref{eq:bellman2}) in the
following Lemma \ref{Lem_bellman2}. Then, we establish the relationship
between the approximate Bellman equation in (\ref{eq:bellman2}) in
Lemma \ref{Lem_bellman2} and the PDE in (\ref{eq:bellman3}) in Theorem
\ref{The_HJB1}.

\emph{1. Relationship between the Equivalent Bellman and Approximate
Bellman Equations}

We establish the following lemma on the approximate Bellman equation
to simplify the equivalent Bellman equation in (\ref{eq:bellman1}):

\begin{lemma}[Approximate Bellman Equation]\label{Lem_bellman2}For
any given weights $\boldsymbol{\beta}$, if
\begin{itemize}
\item there is a unique $\left(\theta^{\ast},\left\{ V^{\ast}\left(\mathbf{Q}\right)\right\} \right)$
that satisfies the Bellman equation and transversality condition in
Theorem \ref{The_opt};
\item there exist $\theta$ and $V\left(\mathbf{Q}\right)$ of class%
\footnote{$f(\mathbf{x})$($\mathbf{x}$ is a $K$-dimensional vector) is of
class $\mathcal{C}^{2}(\mathbb{R}_{+}^{K})$ if the first and second
order partial derivatives of $f(\mathbf{x})$ w.r.t. each element
of $\mathbf{x}$ are continuous when $\mathbf{x}\in\mathbb{R}_{+}^{K}$.%
} $\mathcal{C}^{2}(\mathbb{R}_{+}^{K})$ that solve the following \emph{approximate
Bellman equation}:
\begin{equation}
\begin{aligned} & \theta=\mathbb{E}\bigg[\min_{\boldsymbol{\Omega}\left(\boldsymbol{\chi}\right)}\Big[c\big(\mathbf{Q},\boldsymbol{\Omega}\big(\boldsymbol{\chi}\big)\big)+\sum_{k=1}^{K}\frac{\partial V\left(\mathbf{Q}\right)}{\partial Q_{k}}\Big[\lambda_{k}-R_{k}\big(\mathbf{H},\boldsymbol{\Omega}(\boldsymbol{\chi})\big)\Big]\Big]\bigg|\mathbf{Q}\bigg],\forall\mathbf{Q}\in\boldsymbol{\mathcal{Q}}\end{aligned}
\label{eq:bellman2}
\end{equation}
and for all admissible control policies $\Omega$, the transversality
condition in (\ref{eq:trans1}) is satisfied for $V$,
\end{itemize}
then, we have
\begin{equation}
\theta^{\ast}=\theta+o(1),\quad V^{\ast}\left(\mathbf{Q}\right)=V\left(\mathbf{Q}\right)+o(1),\quad\forall\mathbf{Q}\in\boldsymbol{\mathcal{Q}},
\end{equation}
where the error term $o(1)$ asymptotically goes to zero for sufficiently
small slot duration $\tau$.\hfill\IEEEQED \end{lemma}

\begin{IEEEproof}[Proof of Lemma \ref{Lem_bellman2}]Let $\mathbf{Q}'=(Q_{1}',\cdots,Q_{k}')=\mathbf{Q}(t+1)$
and $\mathbf{Q}=(Q_{1},\cdots,Q_{k})=\mathbf{Q}(t)$. For the queue
dynamics in (\ref{eq:queue}) and sufficiently small $\tau$, we have
$Q_{k}'=Q_{k}-R_{k}\left(\mathbf{H},\mathbf{C}\right)\tau+A_{k}\tau$,
($\forall k$). Therefore, if $V\left(\mathbf{Q}\right)$ is of class
$\mathcal{C}^{2}(\mathbb{R}_{+}^{K})$, we have the following Taylor
expansion on $V\left(\mathbf{Q}'\right)$:
\begin{equation}
\begin{aligned}\mathbb{E}\left[V\left(\mathbf{Q}'\right)\big|\mathbf{Q}\right]= & V\left(\mathbf{Q}\right)+\sum_{k=1}^{K}\frac{\partial V\left(\mathbf{Q}\right)}{\partial Q_{k}}\left(\lambda_{k}-\mathbb{E}\left[R_{k}\big(\mathbf{H},\boldsymbol{\Omega}(\boldsymbol{\chi})\big)\Big|\mathbf{Q}\right]\right)\tau+o(\tau).\end{aligned}
\end{equation}

For notation convenience, let $F_{\boldsymbol{\chi}}(\theta,V,\boldsymbol{\Omega}(\boldsymbol{\chi}))$
denote the \emph{Bellman operator}:
\begin{equation}
\begin{aligned}F_{\boldsymbol{\chi}}(\theta,V,\boldsymbol{\Omega}(\boldsymbol{\chi}))= & \sum_{k=1}^{K}\frac{\partial V\left(\mathbf{Q}\right)}{\partial Q_{k}}\left(\lambda_{k}-R_{k}\big(\mathbf{H},\boldsymbol{\Omega}(\boldsymbol{\chi})\big)\right)-\theta+c\left(\mathbf{Q},\boldsymbol{\Omega}\left(\boldsymbol{\chi}\right)\right)+\nu G_{\boldsymbol{\chi}}\left(V,\boldsymbol{\Omega}\left(\boldsymbol{\chi}\right)\right)\end{aligned}
\end{equation}
for some smooth function $G_{\boldsymbol{\chi}}$ and $\nu=o(1)$
(w.r.t. $\tau$). Denote $F_{\boldsymbol{\chi}}(\theta,V)=\min_{\boldsymbol{\Omega}\left(\mathbf{Q}\right)}F_{\boldsymbol{\chi}}(\theta,V,\boldsymbol{\Omega}(\boldsymbol{\chi}))$.
Suppose $\left(\theta^\ast, V^\ast\right)$ satisfies the Bellman equation in (\ref{eq:bellman1}), we have $\mathbb{E}\left[ F_{\boldsymbol{\chi}}\left( \theta^\ast, V^\ast\right) \big|\mathbf{Q}\right]= \mathbf{0}, \quad \forall \mathbf{Q} \in \boldsymbol{\mathcal{Q}}$. Similarly, if $\left(\theta, V\right)$ satisfies the approximate Bellman equation in (\ref{eq:bellman2}), we have \begin{align} \label{defapp} \mathbb{E}\left[F^\dagger_{\boldsymbol{\chi}}\left( \theta, V\right) \big|\mathbf{Q}\right]= \mathbf{0}, \quad \forall \mathbf{Q} \in \boldsymbol{\mathcal{Q}}, \end{align} where $F^\dagger_{\boldsymbol{\chi}}(\theta, V)=\min_{ \boldsymbol{\Omega}\left( \mathbf{Q} \right)} F^\dagger_{\boldsymbol{\chi}}(\theta, V, \boldsymbol{\Omega}(\boldsymbol{\chi}))$ and $F^\dagger_{\boldsymbol{\chi}}(\theta, V, \boldsymbol{\Omega}(\boldsymbol{\chi}))= F_{\boldsymbol{\chi}}(\theta, V, \boldsymbol{\Omega}(\boldsymbol{\chi}))-\nu G_{\boldsymbol{\chi}}(V,\boldsymbol{\Omega}(\boldsymbol{\chi}))$. We then establish the following lemma.
\begin{lemma}\label{lemma2} If $\left(\theta, V\right)$ satisfies the approximate Bellman equation in (\ref{eq:bellman2}), then $\big|\mathbb{E}\big[F_{\boldsymbol{\chi}}(\theta, V)\big|\mathbf{Q}\big]\big| = o(1)$ for any $\mathbf{Q} \in \boldsymbol{\mathcal{Q}}$.~\hfill\IEEEQED \end{lemma} 
\begin{IEEEproof}[Proof of Lemma~\ref{lemma2}] For any $\boldsymbol{\chi}$, we have $F_{\boldsymbol{\chi}}(\theta, V)=\min_{ \boldsymbol{\Omega}\left( \boldsymbol{\chi} \right)}\big[F^\dagger_{\boldsymbol{\chi}}(\theta, V, \boldsymbol{\Omega}(\boldsymbol{\chi}))+ \nu G_{\boldsymbol{\chi}}(V,\boldsymbol{\Omega}(\boldsymbol{\chi})) \big] \geq \min_{ \boldsymbol{\Omega}\left( \boldsymbol{\chi} \right)} F^\dagger_{\boldsymbol{\chi}}(\theta, V, \boldsymbol{\Omega}(\boldsymbol{\chi}))+ \nu \min_{ \boldsymbol{\Omega}\left( \boldsymbol{\chi} \right)} G_{\boldsymbol{\chi}}(V,\boldsymbol{\Omega}(\boldsymbol{\chi})) $. Besides this, $F_{\boldsymbol{\chi}}(\theta, V)\leq \min_{ \boldsymbol{\Omega}\left( \boldsymbol{\chi} \right)}F^\dagger_{\boldsymbol{\chi}}(\theta, V, \boldsymbol{\Omega}(\boldsymbol{\chi}))+ \nu G_{\boldsymbol{\chi}}(V,\boldsymbol{\Omega}^\dagger (\boldsymbol{\chi} )) $, where $\boldsymbol{\Omega}^\dagger =\arg \min_{ \boldsymbol{\Omega}\left( \boldsymbol{\chi} \right)}F^\dagger_{\boldsymbol{\chi}}(\theta, V, \boldsymbol{\Omega}(\boldsymbol{\chi}))$. Since $\mathbb{E}\big[\min_{ \boldsymbol{\Omega}\left( \boldsymbol{\chi} \right)} F^\dagger_{\boldsymbol{\chi}}(\theta, V, \boldsymbol{\Omega}(\boldsymbol{\chi}))\big|\mathbf{Q}\big]=0$ according to (\ref{defapp}), and $F^\dagger_{\boldsymbol{\chi}}$ and $G_{\boldsymbol{\chi}}$ are all smooth and bounded functions, we have $\big|\mathbb{E}\big[F_{\boldsymbol{\chi}}(\theta, V)\big|\mathbf{Q}\big]\big| = o(1)$ (w.r.t. $\tau$). \end{IEEEproof}
We establish the following lemma to prove Lemma \ref{Lem_bellman2}. 
\begin{lemma} \label{lemma3} Suppose $\mathbf{E}\big[F_{\boldsymbol{\chi}}(\theta^\ast, V^\ast)\big|\mathbf{Q} \big]= 0$ for all $\mathbf{Q}$ together with the transversality condition in (\ref{eq:trans1}) has a unique solution $(\theta^*, V^\ast)$. If $(\theta, V)$ satisfies the approximate Bellman equation in (\ref{eq:bellman2}) and the transversality condition in (\ref{eq:trans1}), then $\theta=\theta^\ast+o\left(1 \right)$, $V\left(\mathbf{Q} \right)=V^\ast\left(\mathbf{Q} \right)+o\left(1 \right)$ for all $\mathbf{Q} $, where $o(1)$ asymptotically goes to zero as $\tau$ goes to zero.~\hfill\IEEEQED \end{lemma} 
\begin{IEEEproof} [Proof of Lemma \ref{lemma3}] Suppose for some $\mathbf{Q}'$, $V\left(\mathbf{Q}' \right)=V^\ast\left(\mathbf{Q}' \right)+\mathcal{O}\left(1 \right)$ (w.r.t. $\tau$). From Lemma~\ref{lemma2}, we have $\big|\mathbb{E}\big[F_{\boldsymbol{\chi}}(\theta, V)\big|\mathbf{Q}\big]\big| = o(1)$ (w.r.t. $\tau$). Letting $\tau \rightarrow 0$, we have $\mathbb{E}\big[F_{\boldsymbol{\chi}}(\theta, V)\big|\mathbf{Q}\big] = 0$ for all $\mathbf{Q}$ and the transversality condition in (\ref{eq:trans1}). However, $V\left(\mathbf{Q}' \right) \neq V^\ast\left(\mathbf{Q}' \right)$ due to $V\left(\mathbf{Q}' \right)=V^\ast\left(\mathbf{Q}' \right)+\mathcal{O}\left(1 \right)$. This contradicts the condition that $(\theta^*, V^\ast)$ is a unique solution of $F_{\boldsymbol{\chi}}(\theta^\ast, V^\ast) = 0$ for all $\mathbf{Q}$ and the transversality condition in (\ref{eq:trans1}). Hence, we must have $V\left(\mathbf{Q} \right)=V^\ast\left(\mathbf{Q} \right)+o\left(1 \right)$ for all $\mathbf{Q} $. Similarly, we can establish $\theta= \theta^\ast + o(1)$. \end{IEEEproof} \end{IEEEproof}

\emph{2. Relationship between the Approximate Bellman Equation and
the PDE}

For notation convenience, we write $J\left(\mathbf{Q}\right)$ in
place of $J\left(\mathbf{Q};\delta\right)$. It can be observed that
if $\left(c^{\infty},\left\{ J\left(\mathbf{Q}\right)\right\} \right)$
satisfies (\ref{eq:bellman3}), it also satisfies (\ref{eq:bellman2}).
Furthermore, since $J\left(\mathbf{Q}\right)=\mathcal{O}(\sum_{k=1}^{K}Q_{k}^{2})$,
then $\lim_{t\rightarrow\infty}\mathbb{E}^{\boldsymbol{\Omega}}\left[J\left(\mathbf{Q}(t)\right)\right]<\infty$
for any admissible policy $\boldsymbol{\Omega}$. Hence, $J\left(\mathbf{Q}\right)=\mathcal{O}(\sum_{k=1}^{K}Q_{k}^{2})$
satisfies the transversality condition in (\ref{eq:trans1}). Next,
we show that the optimal policy $\boldsymbol{\Omega}^{J\ast}$ obtained
from (\ref{eq:bellman3}) is an admissible control policy according
to Definition \ref{Def_adm}.

Define a \emph{Lyapunov function} as $L(\mathbf{Q})=J\left(\mathbf{Q}\right)$.
We define the \emph{conditional queue drift} as $\Delta(\mathbf{Q})=\mathbb{E}^{\boldsymbol{\Omega}^{J\ast}}\big[\sum_{k=1}^{K}\left(Q_{k}(t+1)-Q_{k}(t)\right)\big|\mathbf{Q}(t)=\mathbf{Q}\big]$
and the \emph{conditional Lyapunov drift} as $\Delta L(\mathbf{Q})=\mathbb{E}^{\boldsymbol{\Omega}^{J\ast}}\big[L(\mathbf{Q}(t+1))-L(\mathbf{Q}(t))\big|\mathbf{Q}(t)=\mathbf{Q}\big]$.
We first have the following relationship between $\Delta(\mathbf{Q})$
and $\Delta L(\mathbf{Q})$:
\begin{equation}
\begin{aligned}\Delta L(\mathbf{Q})\geq & \mathbb{E}^{\boldsymbol{\Omega}^{J\ast}}\left[\sum_{k=1}^{K}\frac{\partial L(\mathbf{Q})}{\partial Q_{k}}\left(Q_{k}(t+1)-Q_{k}(t)\right)\bigg|\mathbf{Q}(t)=\mathbf{Q}\right]\overset{(a)}{\geq}\Delta(\mathbf{Q})\end{aligned}
\label{eq:app2Lya1}
\end{equation}
if at least one of $\{Q_{k}:\forall k\}$ is sufficiently large, where
$(a)$ is due to the condition that $\frac{\partial J\left(\mathbf{Q}\right)}{\partial Q_{k}}$
is an increasing function of all $Q_{k}$.

Since $(\lambda_{1},\dots,\lambda_{K})$ is strictly interior to the
stability region $\Lambda$, there exists $\overline{\boldsymbol{\lambda}}=(\lambda_{1}+\kappa_{1},\dots,\lambda_{K}+\kappa_{K})\in\Lambda$
for some positive $\boldsymbol{\kappa}=\{\kappa_{k}:\forall k\}$
\cite{neely}. From \emph{Corollary 1} of \cite{neelyitp}, there
exists a stationary randomized QSI-independent policy $\widetilde{\boldsymbol{\Omega}}$
such that
\begin{equation}
\sum_{k=1}^{K}\mathbb{E}^{\widetilde{\boldsymbol{\Omega}}}\left[\gamma_{k}C_{k}\big|\mathbf{Q}(t)=\mathbf{Q}\right]=\widetilde{C}(\boldsymbol{\kappa})\label{eq:app2Lya2}
\end{equation}
\begin{equation}
\mathbb{E}^{\widetilde{\boldsymbol{\Omega}}}\left[R_{k}(\mathbf{H},\mathbf{C})\big|\mathbf{Q}(t)=\mathbf{Q}\right]\geq\lambda_{k}+\kappa_{k},\quad\forall k,\label{eq:app2Lya3}
\end{equation}
where $\widetilde{C}(\boldsymbol{\kappa})$ is the minimum time-averaging
total fronthaul capacity for the system stability when the arrival
rate is $\overline{\boldsymbol{\lambda}}$. The Lyapunov drift $\Delta L(\mathbf{Q})$
is given by
\begin{equation}
\begin{aligned} & \Delta L(\mathbf{Q})+\mathbb{E}^{\boldsymbol{\Omega}^{J\ast}}\left[\sum_{k=1}^{K}\gamma_{k}C_{k}\tau\bigg|\mathbf{Q}(t)=\mathbf{Q}\right]\\
\approx & \sum_{k=1}^{K}\frac{\partial L(\mathbf{Q})}{\partial Q_{k}}\lambda_{k}\tau+\mathbb{E}^{\boldsymbol{\Omega}^{J\ast}}\left[\sum_{k=1}^{K}\left(\gamma_{k}C_{k}\tau-\frac{\partial L(\mathbf{Q})}{\partial Q_{k}}R_{k}(\mathbf{H},\mathbf{C})\tau\right)\bigg|\mathbf{Q}(t)=\mathbf{Q}\right]\\
\overset{(b)}{\leq} & \sum_{k=1}^{K}\frac{\partial L(\mathbf{Q})}{\partial Q_{k}}\lambda_{k}\tau+\mathbb{E}^{\widetilde{\boldsymbol{\Omega}}}\left[\sum_{k=1}^{K}\left(\gamma_{k}C_{k}\tau-\frac{\partial L(\mathbf{Q})}{\partial Q_{k}}R_{k}(\mathbf{H},\mathbf{C})\tau\right)\bigg|\mathbf{Q}(t)=\mathbf{Q}\right]\\
\overset{(c)}{\leq} & -\sum_{k=1}^{K}\frac{\partial L(\mathbf{Q})}{\partial Q_{k}}\kappa_{k}\tau+\widetilde{C}(\boldsymbol{\kappa})\tau
\end{aligned}
\label{eq:app2Lya4}
\end{equation}
if at least one of $\{Q_{k}:\forall k\}$ is sufficiently large, where
$(b)$ is because $\boldsymbol{\Omega}^{J\ast}$ achieves the minimum
of (\ref{eq:bellman3}) and $(c)$ is due to (\ref{eq:app2Lya2})
and (\ref{eq:app2Lya3}). Combining (\ref{eq:app2Lya4}) with (\ref{eq:app2Lya1}),
we have $\Delta(\mathbf{Q})\leq\Delta L(\mathbf{Q})\leq-\sum_{k=1}^{K}\frac{\partial L(\mathbf{Q})}{\partial Q_{k}}\kappa\tau+\widetilde{C}(\boldsymbol{\kappa})\tau<0$
if at least one of $\{Q_{k}:\forall k\}$ is sufficiently large. Therefore,\textcolor{red}{{}
${\normalcolor \mathbb{E}\big[A_{k}-G_{k}(\mathbf{H},\boldsymbol{\Omega}^{J\ast}(\boldsymbol{\chi}))\big|\mathbf{Q}\big]<0}$}
when $Q_{k}>\overline{Q}_{k}$ for some large $\overline{Q}_{k}$.
Let $\phi_k(r,\mathbf{Q})=\ln \big(\mathbb{E}\big[e^{\left(A_k-G_k(\mathbf{H},\boldsymbol{\Omega}^{J \ast}(\boldsymbol{\chi}) ) \right)r}\big| \mathbf{Q} \big] \big)$ be the \emph{semi-invariant moment generating function} of $A_k-G_k\big(\mathbf{H},\boldsymbol{\Omega}^{J \ast}(\boldsymbol{\chi})\big)$. Then, $\phi_k(r,\mathbf{Q})$ will have a unique positive root $r_k^\ast(\mathbf{Q})$ ($\phi_k(r_k^\ast(\mathbf{Q}),\mathbf{Q})=0$) \cite{dspgal}. Let $r_k^\ast= r_k^\ast(\overline{\mathbf{Q}})$, where $\overline{\mathbf{Q}}=(\overline{Q}_1, \dots, \overline{Q}_K)$. Using the Kingman bound \cite{dspgal} result that $F_k(x) \triangleq \Pr\big[ Q_k \geq x \big] \leq e^{-r_k^\ast x} $, if $x \geq \overline{x}_k$ for sufficiently large $\overline{x}_k$, we have \begin{align} &\mathbb{E}^{\boldsymbol{\Omega}^{J \ast}} \left[J\left(\mathbf{Q}\right) \right] \nonumber\\ \leq& C \sum_{k=1}^K \mathbb{E}^{\boldsymbol{\Omega}^{J \ast}} \left[ Q_k^2 \right]=C\sum_{k=1}^K \left[\int_0^{\infty} \Pr \left[Q_k^2 >s \right] \mathrm{d}s \right] \notag \\ \leq & C \sum_{k=1}^K \left[ \int_{0}^{\overline{x}_k^2}F_k(s^{1/2}) \mathrm{d}s + \int_{\overline{x}_k^2}^{\infty} F_k(s^{1/2})\mathrm{d}s \right] \nonumber\\ \leq& C \sum_{k=1}^K \left[\overline{x}_k^2+ \int_{\overline{x}_k^2}^{\infty} e^{-r_k^\ast s^{1/2}} \mathrm{d}s \right] < \infty \end{align} for some constant $C$. Therefore, $\boldsymbol{\Omega}^{J \ast}$ is an admissible control policy and we have $V \left(\mathbf{Q} \right)=J\left(\mathbf{Q}\right)$ and $\theta=c^\infty$.

Combining Lemma \ref{Lem_bellman2}, we have $V^{\ast}\left(\mathbf{Q}\right)=J\left(\mathbf{Q}\right)+o(1)$
and $\theta^{\ast}=c^{\infty}+o(1)$ for sufficiently small $\tau$.

\section{Proof of Lemma \ref{Lem_weak}}

The coupling among the $K$ uplink data queues is induced by $\mathbf{S}(t)$
in the expression of $R_{k}$ in (\ref{eq:rate}). According to Assumption
\ref{Ass_ZF}, $\mathbf{S}(t)=\mathbf{H}(t)^{-1}$. The time index
$t$ is omitted in this proof for simplicity of expression. We adopt
the adjoint matrix to obtain the inverse of the channel matrix $\mathbf{H}$
as
\begin{equation}
\mathbf{S}=\frac{1}{\det(\mathbf{H})}\mathrm{adj}(\mathbf{H})=\frac{1}{\left|\mathbf{H}\right|}\left[\begin{array}{cccc}
M_{11} & M_{12} & \cdots & M_{1K}\\
M_{21} & M_{22} & \cdots & M_{2K}\\
\vdots & \vdots & \ddots & \vdots\\
M_{K1} & M_{K2} & \cdots & M_{KK}
\end{array}\right]^{T},
\end{equation}
where $M_{kj}$ is the $(k,j)$ algebraic cofactor, which is the determinant
of the submatrix formed by deleting the $k$-th row and $j$-th column
of $\mathbf{H}$ multiplied by $(-1)^{k+j}$.

With $\delta=\max\left\{ L_{kj}:\forall k\neq j\right\} $, we can
rewrite the channel matrix $\mathbf{H}$ as
\begin{equation}
\left[\begin{array}{cccc}
\mathcal{O}(1) & \mathcal{O}(\sqrt{\delta}) & \cdots & \mathcal{O}(\sqrt{\delta})\\
\mathcal{O}(\sqrt{\delta}) & \mathcal{O}(1) & \cdots & \mathcal{O}(\sqrt{\delta})\\
\vdots & \vdots & \ddots & \vdots\\
\mathcal{O}(\sqrt{\delta}) & \mathcal{O}(\sqrt{\delta}) & \cdots & \mathcal{O}(1)
\end{array}\right],
\end{equation}
where the $K$ diagonal entries are $\mathcal{O}(1)$ and the other
entries are $\mathcal{O}(\sqrt{\delta})$.

If $k\neq j$, the submatrix formed by deleting the $k$-th row and
$j$-th column of $\mathbf{H}$ includes $K-2$ diagonal entries of
$\mathbf{H}$, i.e., $\mathcal{O}(1)$. As a result, when calculating
the determinant of the submatrix, each term of the determinant is
the product of $K-1$ entries and at least one $\mathcal{O}(\sqrt{\delta})$
term is included. Therefore, we obtain the coupling intensity $||S_{kj}(t)||^{2}=\mathcal{O}\left(\delta\right),\forall k\neq j$,
and Lemma \ref{Lem_weak} holds.

\section{Proof of Lemma \ref{Lem_base}}

We first prove that $J\left(\mathbf{Q};0\right)=\sum_{k=1}^{K}J_{k}\left(Q_{k}\right)$.
The PDE in (\ref{eq:bellman3}) for the base system is
\begin{equation}
\begin{aligned} & \mathbb{E}\bigg[\min_{\boldsymbol{\Omega}(\boldsymbol{\chi})}\bigg[\sum_{k=1}^{K}\bigg(\beta_{k}\frac{Q_{k}}{\lambda_{k}}+\gamma_{k}C_{k}+\frac{\partial J\left(\mathbf{Q};0\right)}{\partial Q_{k}}\Big(\lambda_{k}-R_{k}\big(\mathbf{H},\mathbf{C}\big)\Big)\bigg)\bigg]\bigg|\mathbf{Q}\bigg]-c^{\infty}=0.\end{aligned}
\label{eq:app4HJB}
\end{equation}
We have the following lemma to prove the decomposable structures of
$J\left(\mathbf{Q};0\right)$ and $c^{\infty}$ in (\ref{eq:app4HJB}).

\begin{lemma} [Decomposed Optimality Equation]\label{Lem_decom}Suppose
there exist $c_{k}^{\infty}$ and $J_{k}\left(Q_{k}\right)\in\mathbb{C}^{2}\left(\mathbb{R}_{+}\right)$
that solve the following per-flow optimality equation (PFOE):
\begin{equation}
\begin{aligned} & \mathbb{E}\bigg[\min_{C_{k}\geq0}\bigg[\beta_{k}\frac{Q_{k}}{\lambda_{k}}+\gamma_{k}C_{k}+J_{k}'(Q_{k})\Big(\lambda_{k}-R_{k}^{0}\big(H_{kk},C_{k}\big)\Big)\bigg]\bigg|Q_{k}\bigg]-c_{k}^{\infty}=0,\end{aligned}
\label{eq:app4perflow}
\end{equation}
where $R_{k}^{0}\big(H_{kk},C_{k}\big)=\log_{2}\left(1+\frac{P\left\Vert H_{kk}\right\Vert ^{2}}{N_{0}+N_{k}}\right)$
and $N_{k}=\frac{P\left\Vert H_{kk}\right\Vert ^{2}+N_{0}}{2^{C_{k}}-1}$.
Then, $J\left(\mathbf{Q};0\right)=\sum_{k=1}^{K}J_{k}\left(Q_{k}\right)$
and $c^{\infty}=\sum_{k=1}^{K}c_{k}^{\infty}$ satisfy (\ref{eq:app4HJB}).\hfill\IEEEQED \end{lemma}

Lemma \ref{Lem_decom} can be proved using the fact that the dynamics
of the $K$ queues at the UEs are decoupled when $\delta=0$. The
details are omitted for conciseness.

Next, we solve the optimization problem in (\ref{eq:app4perflow}).
The optimal fronthaul capacity $C_{k}^{*}$ from (\ref{eq:app4perflow})
is given by
\begin{equation}
C_{k}^{*}=\left(\log_{2}\left(\frac{P\left\Vert H_{kk}\right\Vert ^{2}}{N_{0}}\left(\frac{J_{k}'(Q_{k})}{\gamma_{k}}-1\right)^{+}\right)\right)^{+}.
\end{equation}
Substituting the optimal allocated fronthaul capacity $C_{k}^{\ast}$
into (\ref{eq:app4perflow}), and using the fact that $\left\Vert H_{kk}\right\Vert ^{2}$
follows a negative exponential distribution with mean $L_{kk}$ according
to Assumption \ref{Ass_csi}, we calculate the expectations in (\ref{eq:app4perflow})
as follows:

If $J_{k}'(Q_{k})>\gamma_{k}$, the expected fronthaul capacity is
\begin{equation}
\begin{aligned}\mathbb{E}\left[\gamma_{k}C_{k}^{\ast}\big|Q_{k}\right]= & \int_{\frac{N_{0}\gamma_{k}}{PL_{kk}\left(J_{k}'(Q_{k})-\gamma_{k}\right)}}^{\infty}\log_{2}\left(\frac{PL_{kk}x}{N_{0}}\left(\frac{J_{k}'(Q_{k})}{\gamma_{k}}-1\right)\right)e^{-x}dx\\
= & \frac{\gamma_{k}}{\ln2}E_{1}\left(\frac{N_{0}\gamma_{k}}{PL_{kk}\left(J_{k}'(Q_{k})-\gamma_{k}\right)}\right),
\end{aligned}
\label{eq:app4exp1}
\end{equation}
where $E_{1}(z)\triangleq\int_{z}^{\infty}\frac{e^{-t}}{t}\mathrm{d}t$
is the exponential integral function. Otherwise, $\mathbb{E}\left[\gamma_{k}C_{k}^{\ast}\big|Q_{k}\right]=0$.
Similarly, if $J_{k}'(Q_{k})>\gamma_{k}$, the expected data rate
is 
\begin{equation}
\begin{aligned}\mathbb{E}\left[R_{k}^{0}\big(H_{kk},C_{k}^{*}\big)\big|Q_{k}\right]= & \int_{\frac{N_{0}\gamma_{k}}{PL_{kk}\left(J_{k}'(Q_{k})-\gamma_{k}\right)}}^{\infty}\log_{2}\left(\frac{1+PL_{kk}x/N_{0}}{1+\frac{1}{\left(J_{k}'(Q_{k})/\gamma_{k}-1\right)}}\right)e^{-x}dx\\
= & \frac{e^{\frac{N_{0}}{PL_{kk}}}}{\ln2}E_{1}\left(\frac{N_{0}J_{k}'(Q_{k})}{PL_{kk}\left(J_{k}'(Q_{k})-\gamma_{k}\right)}\right).
\end{aligned}
\label{eq:app4exp2}
\end{equation}
Otherwise, $\mathbb{E}\left[R_{k}^{0}\big(H_{kk},C_{k}^{*}\big)\big|Q_{k}\right]=0$. 

We then calculate $c_{k}^{\infty}$. Since (\ref{eq:app4perflow})
should hold when $Q_{k}=0$, we have
\begin{equation}
c_{k}^{\infty}=\mathbb{E}\left[\gamma_{k}C_{k}^{\ast}\big|Q_{k}=0\right]\label{eq:app4exp3}
\end{equation}
\begin{equation}
\mathbb{E}\left[R_{k}^{0}\big(H_{kk},C_{k}\big)\big|Q_{k}=0\right]=\lambda_{k}.\label{eq:app4exp4}
\end{equation}
Using (\ref{eq:app4exp1}) and (\ref{eq:app4exp2}), we can calculate
$c_{k}^{\infty}$ as shown in Lemma \ref{Lem_base}. Substituting
(\ref{eq:app4exp1}), (\ref{eq:app4exp2}), and $c_{k}^{\infty}$
into (\ref{eq:app4perflow}) and letting $a_{k}\triangleq\frac{N_{0}}{PL_{kk}}$,
we have the following ODE:
\begin{equation}
\begin{aligned} & \beta_{k}\frac{Q_{k}}{\lambda_{k}}+\frac{\gamma_{k}}{\ln2}E_{1}\left(\frac{a_{k}\gamma_{k}}{J_{k}'(Q_{k})-\gamma_{k}}\right)+J_{k}'\left(Q_{k}\right)\lambda_{k}-J_{k}'\left(Q_{k}\right)\frac{e^{a_{k}}}{\ln2}E_{1}\left(\frac{a_{k}J_{k}'(Q_{k})}{J_{k}'(Q_{k})-\gamma_{k}}\right)-c_{k}^{\infty}=0.\end{aligned}
\label{eq:perflowODE}
\end{equation}

According to Section 0.1.7.3 of \cite{ODE}, we can obtain the parametric
solution of (\ref{eq:perflowODE}), as shown in (\ref{eq:perflow})
in Lemma \ref{Lem_base}.

\section{Proof of Theorem \ref{The_first}}

Taking the first order Taylor expansion of the L.H.S. of the Bellman
equation in (\ref{eq:bellman3}) at $L_{ij}=0$ ($\forall i\neq j$),
$C_{k}=C_{k}^{\ast}$, where $C_{k}^{\ast}$ minimizes the L.H.S.
of (\ref{eq:app4perflow}), and using parametric optimization analysis
\cite{perturbation}, we have the following result regarding the approximation
error:
\begin{equation}
J\left(\mathbf{Q};\delta\right)-J\left(\mathbf{Q};0\right)=\sum_{i=1}^{K}\sum_{j=1,j\neq i}^{K}L_{ij}\widetilde{J}_{ij}(\mathbf{Q})+\mathcal{O}(\delta{}^{2}),\label{eq:app5taylor}
\end{equation}
where $\widetilde{J}_{ij}(\mathbf{Q})$ captures the coupling terms
in $J\left(\mathbf{Q}\right)$ satisfying
\begin{equation}
\begin{aligned} & \sum_{k=1}^{K}\left(\lambda_{k}-\mathbb{E}\left[\left.\log_{2}\left(1+\frac{PL_{kk}\left\Vert \widetilde{H}_{kk}\right\Vert ^{2}}{N_{0}+N_{k}^{*}}\right)\right|\mathbf{Q}\right]\right)\frac{\partial\widetilde{J}_{ij}\left(\mathbf{Q}\right)}{\partial Q_{k}}\\
 & +\mathbb{E}\left[\frac{\left\Vert \widetilde{H}_{ij}\right\Vert ^{2}}{\ln2}\left(\frac{\frac{J_{i}'\left(Q_{i}\right)}{N_{0}+N_{i}^{*}}}{1+\frac{N_{0}+N_{i}^{*}}{PL_{ii}\left\Vert \widetilde{H}_{ii}\right\Vert ^{2}}}\sum\limits _{l=1,l\neq i}^{K}\frac{N_{0}+N_{l}^{*}}{L_{ll}\left\Vert \widetilde{H}_{ll}\right\Vert ^{2}}\right.\right.\\
 & \left.\left.\left.+\sum\limits _{k=1,k\neq i,j}^{K}\frac{\frac{J_{k}'\left(Q_{k}\right)}{N_{0}+N_{k}^{*}}}{\left(1+\frac{N_{0}+N_{k}^{*}}{PL_{kk}\left\Vert \widetilde{H}_{kk}\right\Vert ^{2}}\right)}\frac{N_{0}+N_{j}^{*}}{L_{jj}\left\Vert \widetilde{H}_{jj}\right\Vert ^{2}}\right)\right|\mathbf{Q}\right]=\widetilde{\theta}_{ij},
\end{aligned}
\label{eq:app5pde}
\end{equation}
with boundary condition $\widetilde{J}_{ij}\left(\mathbf{Q}\right)\big|_{Q_{i}=0}=0$
or $\widetilde{J}_{ij}\left(\mathbf{Q}\right)\big|_{Q_{j}=0}=0$,
where $N_{k}^{*}=\frac{PL_{kk}\left\Vert \widetilde{H}_{kk}\right\Vert ^{2}+N_{0}}{2^{C_{k}^{*}}-1}$
and $\widetilde{\theta}_{ij}=\frac{\partial\theta}{\partial L_{ij}}$
is constant (where we treat $\theta$ as a function of $\left\{ L_{ij}:\forall i\neq j\right\} $).
According to (\ref{eq:app4exp1}), we have
\begin{equation}
\mathbb{E}\left[\left.\log_{2}\left(1+\frac{PL_{kk}\left\Vert \widetilde{H}_{kk}\right\Vert ^{2}}{N_{0}+N_{k}^{*}}\right)\right|\mathbf{Q}\right]=\frac{e^{a_{k}}E_{1}\left(a_{k}\right)}{\ln2}\mathcal{O}\left(1\right).
\end{equation}
Then, we calculate the second term in (\ref{eq:app5pde}) and each
part is calculated as follows: 
\begin{equation}
\begin{aligned}\mathbb{E}\left[\left.\frac{\frac{J_{i}'\left(Q_{i}\right)}{N_{0}+N_{i}^{*}}}{1+\frac{N_{0}+N_{i}^{*}}{PL_{ii}\left\Vert \widetilde{H}_{ii}\right\Vert ^{2}}}\right|\mathbf{Q}\right]= & \mathbb{E}\left[\left.\frac{PL_{ii}\left\Vert \widetilde{H}_{ii}\right\Vert ^{2}}{\left(N_{0}+PL_{ii}\left\Vert \widetilde{H}_{ii}\right\Vert ^{2}\right)N_{0}}\right|\mathbf{Q}\right]\mathcal{O}\left(J_{i}'(Q_{i})\right)\\
= & \frac{\beta_{i}}{\lambda_{i}}\left(1-\frac{a_{i}e^{a_{i}}E_{1}\left(a_{i}\right)}{N_{0}}\right)\mathcal{O}\left(Q_{i}\right)
\end{aligned}
\end{equation}
\begin{equation}
\begin{aligned}\mathbb{E}\left[\left.\frac{N_{0}+N_{j}^{*}}{L_{jj}\left\Vert \widetilde{H}_{jj}\right\Vert ^{2}}\right|\mathbf{Q}\right]= & \frac{2N_{0}}{L_{jj}}\mathbb{E}\left[\left.\frac{1}{\left\Vert \widetilde{H}_{jj}\right\Vert ^{2}}\right|\mathbf{Q}\right]\mathcal{O}\left(1\right)=\frac{2N_{0}}{L_{jj}}\mathcal{O}\left(1\right).\end{aligned}
\end{equation}
Substituting these calculation results into (\ref{eq:app5pde}), we
rewrite the PDE as
\begin{equation}
\begin{aligned} & \sum_{k=1}^{K}\left(\lambda_{k}-\frac{e^{a_{k}}E_{1}\left(a_{k}\right)}{\ln2}\mathcal{O}\left(1\right)\right)\frac{\partial\widetilde{J}_{ij}\left(\mathbf{Q}\right)}{\partial Q_{k}}\\
 & +\frac{\beta_{i}}{\lambda_{i}\ln2}\left(1-\frac{a_{i}e^{a_{i}}E_{1}\left(a_{i}\right)}{N_{0}}\right)\sum\limits _{l=1,l\neq i}^{K}\frac{2N_{0}}{L_{ll}}\mathcal{O}\left(Q_{i}\right)\\
 & +\sum\limits _{k=1,k\neq i,j}^{K}\frac{\beta_{k}}{\lambda_{k}\ln2}\left(1-\frac{a_{k}e^{a_{k}}E_{1}\left(a_{k}\right)}{N_{0}}\right)\frac{2N_{0}}{L_{jj}}\mathcal{O}\left(Q_{k}\right)=0.
\end{aligned}
\end{equation}
Using \emph{3.8.2.1} of \cite{PDE} and taking into account the boundary
conditions, we obtain that 
\begin{equation}
\begin{aligned}\widetilde{J}_{ij}\left(\mathbf{Q}\right)= & \frac{\frac{\beta_{i}}{\lambda_{i}}\left(1-\frac{a_{i}e^{a_{i}}E_{1}\left(a_{i}\right)}{N_{0}}\right)\sum\limits _{l=1,l\neq i}^{K}\frac{N_{0}}{L_{ll}}}{e^{a_{i}}E_{1}\left(a_{i}\right)-\lambda_{i}\ln2}\mathcal{O}\left(Q_{i}^{2}\right)+\frac{N_{0}}{L_{jj}}\sum\limits _{k=1,k\neq i,j}^{K}\frac{\frac{\beta_{k}}{\lambda_{k}}\left(1-\frac{a_{k}e^{a_{k}}E_{1}\left(a_{k}\right)}{N_{0}}\right)}{e^{a_{k}}E_{1}\left(a_{k}\right)-\lambda_{k}\ln2}\mathcal{O}\left(Q_{k}^{2}\right).\end{aligned}
\end{equation}
Substituting it into (\ref{eq:app5taylor}) and exchanging the indices
$i$ and $k$, we obtain the first order perturbation (\ref{eq:firstorder})
in Theorem \ref{The_first}.

\section{Proof of Lemma \ref{The:convex}}

We adopt the following argument to prove the convexity \cite{Boyd}:
given two feasible points $\mathbf{x}_{1}$ and $\mathbf{x}_{2}$,
define $g(t)=f(t\mathbf{x}_{1}+(1-t)\mathbf{x}_{2})$, $0\leq t\leq1$,
then $f(\mathbf{x})$ is a convex function of $\mathbf{x}$ if and
only if $g(t)$ is a convex function of $t$, which is equivalent
to $\frac{\mathrm{d}^{2}g(t)}{\mathrm{d}t^{2}}\geq0$ for $0\leq t\leq1$.
To use this argument, we rewrite problem (\ref{eq:utility}) as
\begin{equation}
\min_{\mathbf{C}}\ f\left(\mathbf{C},\delta\right)=\sum_{k=1}^{K}\left(\gamma_{k}C_{k}-\frac{\partial\widetilde{V}\left(\mathbf{Q}\right)}{\partial Q_{k}}R_{k}\left(\mathbf{H},\mathbf{C}\right)\right).
\end{equation}

Consider the convex combination of two feasible solutions, $\mathbf{C}^{(1)}=\big\{ C_{k}^{(1)}:\forall k\big\}$
and $\mathbf{C}^{(2)}=\big\{ C_{k}^{(2)}:\forall k\big\}\big\}$,
as $\mathbf{C}^{c}=\big\{ C_{k}^{c}=tC_{k}^{(1)}+(1-t)C_{k}^{(2)}:\forall k\big\}$
and $0\leq t\leq1$. When $\delta$ is sufficiently small, the second
order derivative of $f\left(\mathbf{C}^{c},\delta\right)$ is calculated
as
\begin{equation}
\begin{aligned}\frac{\mathrm{d}^{2}f\left(\mathbf{C}^{c},\delta\right)}{\mathrm{d}t^{2}}= & \sum_{k=1}^{K}\Bigg(\frac{\partial\widetilde{V}\left(\mathbf{Q}\right)}{\partial Q_{k}}\ln2\Bigg(\frac{-P^{2}-2PZ_{k}}{\left(P+Z_{k}\right)^{2}Z_{k}^{2}}\left(\sum_{j=1}^{K}\frac{X_{j}Y_{j}\left\Vert S_{kj}\right\Vert ^{2}(C_{j}^{(1)}-C_{j}^{(2)})}{(X_{j}-1)^{2}}\right)^{2}\\
 & +\frac{P}{\left(P+Z_{k}\right)Z_{k}}\sum_{j=1}^{K}\frac{\left(X_{j}^{2}+X_{j}\right)Y_{j}\left\Vert S_{kj}\right\Vert ^{2}(C_{j}^{(1)}-C_{j}^{(2)})^{2}}{(X_{j}-1)^{3}}\Bigg)\Bigg),
\end{aligned}
\end{equation}
where $X_{j}=2^{tC_{j}^{(1)}+(1-t)C_{j}^{(2)}}$ and $Z_{k}=\sum_{j=1}^{K}\left\Vert S_{kj}\right\Vert ^{2}\left(N_{0}+N_{j}\right)$.
When $\delta$ is sufficiently small, the terms in the order of $\mathcal{O}(\delta)$
can be ignored and we simplify $\frac{\mathrm{d}^{2}f\left(\mathbf{C}^{c},\delta\right)}{\mathrm{d}t^{2}}$
as 
\begin{equation}
\begin{aligned}\frac{\mathrm{d}^{2}f\left(\mathbf{C}^{c},\delta\right)}{\mathrm{d}t^{2}}\approx & \sum_{k=1}^{K}\frac{\frac{\partial\widetilde{V}\left(\mathbf{Q}\right)}{\partial Q_{k}}PX_{k}^{3}Y_{k}\left(PN_{0}+\left\Vert S_{kk}\right\Vert ^{2}N^{2}\right)\left\Vert S_{kk}\right\Vert ^{4}\left(C_{k}^{(1)}-C_{k}^{(2)}\right)^{2}\ln2}{(X_{k}-1)^{4}\left(P+Z_{k}\right)^{2}Z_{k}^{2}}.\end{aligned}
\label{eq:app7}
\end{equation}

It is obvious that $\frac{\mathrm{d}^{2}f\left(\mathbf{C}^{c},\delta\right)}{\mathrm{d}t^{2}}>0$
in (\ref{eq:app7}), and thus, the problem (\ref{eq:utility}) is
a convex optimization problem for sufficiently small $\delta$.

\section{Proof of Theorem \ref{The:convergence}}

We prove the convergence by the fictitious game model \cite{Huang}.
We first construct the following capacity-price fictitious game model.
The optimization problem of the fictitious capacity player $k$ is
\begin{equation}
\max_{C_{k}}\quad u_{k}^{FW}=\frac{\partial\widetilde{V}\left(\mathbf{Q}\right)}{\partial Q_{k}}R_{k}\left(\mathbf{H},\mathbf{C}\right)-\gamma_{k}C_{k}+\sum_{i=1,i\neq k}\pi_{ik}C_{k}.
\end{equation}
The optimization problem of the fictitious price player is
\begin{equation}
\max_{\pi{}_{ik}}\quad u_{ik}^{FC}=-\left(\pi_{ik}-\frac{\partial\left(\frac{\partial\widetilde{V}\left(\mathbf{Q}\right)}{\partial Q_{i}}R_{i}\left(\mathbf{H},\mathbf{C}\right)\right)}{\partial C_{k}}\right)^{2}.
\end{equation}

Each player in this game adopts the myopic best response (MBR) to
update his strategy. From \cite{super}, the MBR updates converge
to Nash Equilibrium in the supermodular games, in which the payoff
function is supermodular in player $i$'s strategy and has increasing
differences between any component of player $k$'s strategy and any
component of any other player's strategy. Now, we check if this fictitious
game is supermodular. It is obvious that each player's payoff function
is supermodular in its own one-dimensional strategy. According to
the method in \cite{Huang}, we have $\frac{\partial u_{k}^{FW}}{\partial C_{k}\partial\pi{}_{ik}}=1>0,\forall i\neq k$
and the increasing difference condition is satisfied. Therefore, the
fictitious game is a supermodular game and always converges. 

When $\delta$ is sufficiently small, according to Lemma \ref{The:convex},
the problem is convex and the supermodular game converges to the unique
global optimal solution of the per-stage problem. Furthermore, the
approximation error of the priority function in Theorem \ref{The_first}
approaches 0 with sufficiently small $\delta$. Therefore, the supermodular
game converges to the optimal solution of Problem \ref{Pro_MDP} with
sufficiently small $\delta$.

\end{document}